\documentclass[fleqn,usenatbib]{mnras}

\DeclareRobustCommand{\VAN}[3]{#2}
\let\VANthebibliography\thebibliography
\def\thebibliography{\DeclareRobustCommand{\VAN}[3]{##3}\VANthebibliography}

\usepackage{graphicx}	
\usepackage{amsmath}	
\usepackage{amssymb}	

\usepackage{newtxtext,newtxmath}

\usepackage[T1]{fontenc}

\usepackage[usenames,dvipsnames]{xcolor}

\newcommand{\lya}{Ly$\alpha$}
\newcommand{\lyb}{Ly$\beta$}

\newcommand{\hi}{\ion{H}{i}}
\newcommand{\hii}{\ion{H}{ii}}
\newcommand{\oi}{\ion{O}{i}}
\newcommand{\oiii}{\ion{O}{iii}}
\newcommand{\siii}{\ion{Si}{ii}}
\newcommand{\siiv}{\ion{Si}{iv}}
\newcommand{\civ}{\ion{C}{iv}}
\newcommand{\nv}{\ion{N}{v}}
\newcommand{\cii}{\ion{C}{ii}}
\newcommand{\kms}{km~s$^{-1}$}

\newcommand{\hinvMpc}{$h^{-1}\,$Mpc}

\newcommand{\meanflux}{$\langle F \rangle$}
\newcommand{\meandiff}{$\langle \Delta F \rangle$}
\newcommand{\Lquasar}{$L_{\rm q}$}

\title[Damping wing from a $z < 6$ \lya\ trough]{Damping wing absorption associated with a giant \lya\ trough at $z < 6$: direct evidence for late-ending reionization}

\author[G. D. Becker et al.]{George D. Becker,$^1$\thanks{E-mail: george.becker@ucr.edu} James S. Bolton,$^2$ Yongda Zhu,$^{1,3}$ and Seyedazim Hashemi$^{1}$
\\
$^1$Department of Physics \& Astronomy, University of California, Riverside, CA, 92521, USA \\
$^2$School of Physics and Astronomy, University of Nottingham, University Park, Nottingham, NG7 2RD, UK \\
$^3$Steward Observatory, University of Arizona, 933 N Cherry Ave, Tucson, AZ 85721, USA
}

\date{Accepted XXX. Received YYY; in original form ZZZ}

\pubyear{2023}

\begin{document}
\label{firstpage}
\pagerange{\pageref{firstpage}--\pageref{lastpage}}
\maketitle

\begin{abstract}

Multiple observations now suggest that the hydrogen reionization may have ended well below redshift six.  While there has previously been no conclusive proof of extended neutral islands in the $z < 6$ intergalactic medium, it is possible that such islands give rise to the giant \lya\ absorption troughs seen in the spectra of high-redshift quasars.  Here we present evidence that the deepest and longest-known \lya\ trough at $z < 6$, towards ULAS J0148+0600 (J0148), is associated with damping wing absorption.  The evidence comes from a window of strong \lya\ transmission at the edge of the J0148 proximity zone.  We show that the relatively smooth profile of this transmission window is highly unlikely to arise from resonant absorption alone, but  is consistent with the presence of a damping wing.  We further argue that the damping wing is unlikely to arise from a compact source due to the lack of associated metal lines, and is more likely to arise from an extended neutral island associated with the giant \lya\ trough.  We investigate the physical conditions that may give rise to the strong transmission window, and speculate that it may signal an usually deep void, nearby ionizing sources, and/or the recent passage of an ionization front.

\end{abstract}


\section{Introduction}

The reionization of the intergalactic medium (IGM) was a milestone in cosmic history.  As a global baryonic phase change, it represents one of the most significant interactions between galaxies and their environments.  Determining when and how reionization occurred is therefore essential for developing a self-consistent picture of galaxy evolution and IGM physics in the early Universe \citep[for recent reviews, see][]{dayal2018,wise2019,robertson2022,choudhury2022,fan2023}.

Over the past several years, a consensus has begun to emerge that reionization may have ended significantly later than previously believed.   \lya\ transmission in the spectra of the first known $z \sim 6$ quasars signaled that reionization was at least largely complete by that redshift \citep[e.g.,][]{fan2006a}.  With larger samples and higher-quality spectra, however, it has become clear that the high degree of scatter in the large-scale \lya\ forest transmission is not consistent with a fully reionized IGM, at least not one with a uniform ionizing UV background and temperature-density relation \citep{becker2015,daloisio2015,daloisio2018,davies2016,bosman2018,bosman2022,eilers2018,yang2020a}.  Models wherein reionization ended closer to $z \sim 5$, however, can naturally reproduce this scatter while remaining consistent with other constraints on reionization such as the electron optical depth to the cosmic microwave background and the apparent attenuation of \lya\ emission from quasars and galaxies at $z \gtrsim 7$ \citep{kulkarni2019,keating2020a,keating2020,nasir2020,choudhury2021,qin2021}.  
A strong indication that reionization was still underway at $z \sim 6$ comes from the rapid evolution of the mean free path of ionizing photons, which has been measured directly from the Lyman continuum transmission of quasar spectra \citep{becker2021,bosman2021,zhu2023} and indirectly from large-scale \lya\ opacity fluctuations \citep{gaikwad2023,davies2023a}.  The increase in the mean free path from $z \sim 6$ to 5 is more rapid than would be expected for a fully reionized IGM \citep{daloisio2020}, but is broadly consistent with the photoevaporation of ionizing sinks at the end of reionization \citep{keating2020a,cain2021,garaldi2022,lewis2022}.  

While a late-ending reionization scenario is supported by a number of constraints, obtaining direct observational evidence of neutral gas in the $z < 6$ IGM has been challenging due to the fact that resonant \lya\ absorption saturates for even very small hydrogen neutral fractions ($x_{\rm H\, I} \gtrsim 10^{-4}$) .  The best potential tracers of neutral regions are the ``dark gaps'' in the \lya\ and \lyb\ forests \citep[e.g.,][]{djorgovski2001,becker2001,songaila2002,fan2006a,zhu2021,zhu2022}.  These are extended ($\gtrsim$10~\hinvMpc) regions of nearly complete absorption, which in multiple cases have been found to trace low-density regions based on the number density of surrounding galaxies \citep{becker2018,kashino2020,christenson2021,ishimoto2022}.  Dark gaps are naturally explained as absorption by the last remaining neutral islands at the end of reionization \citep[e.g.,][]{keating2020a,nasir2020}.  Observationally, however, it is difficult to distinguish neutral islands from ionized regions of high opacity that could arise from large fluctuations in the ionizing UV background \citep{davies2016,nasir2020}.

\begin{figure*}
   \centering
   \begin{minipage}{\textwidth}
   \begin{center}
   \includegraphics[width=1.0\textwidth]{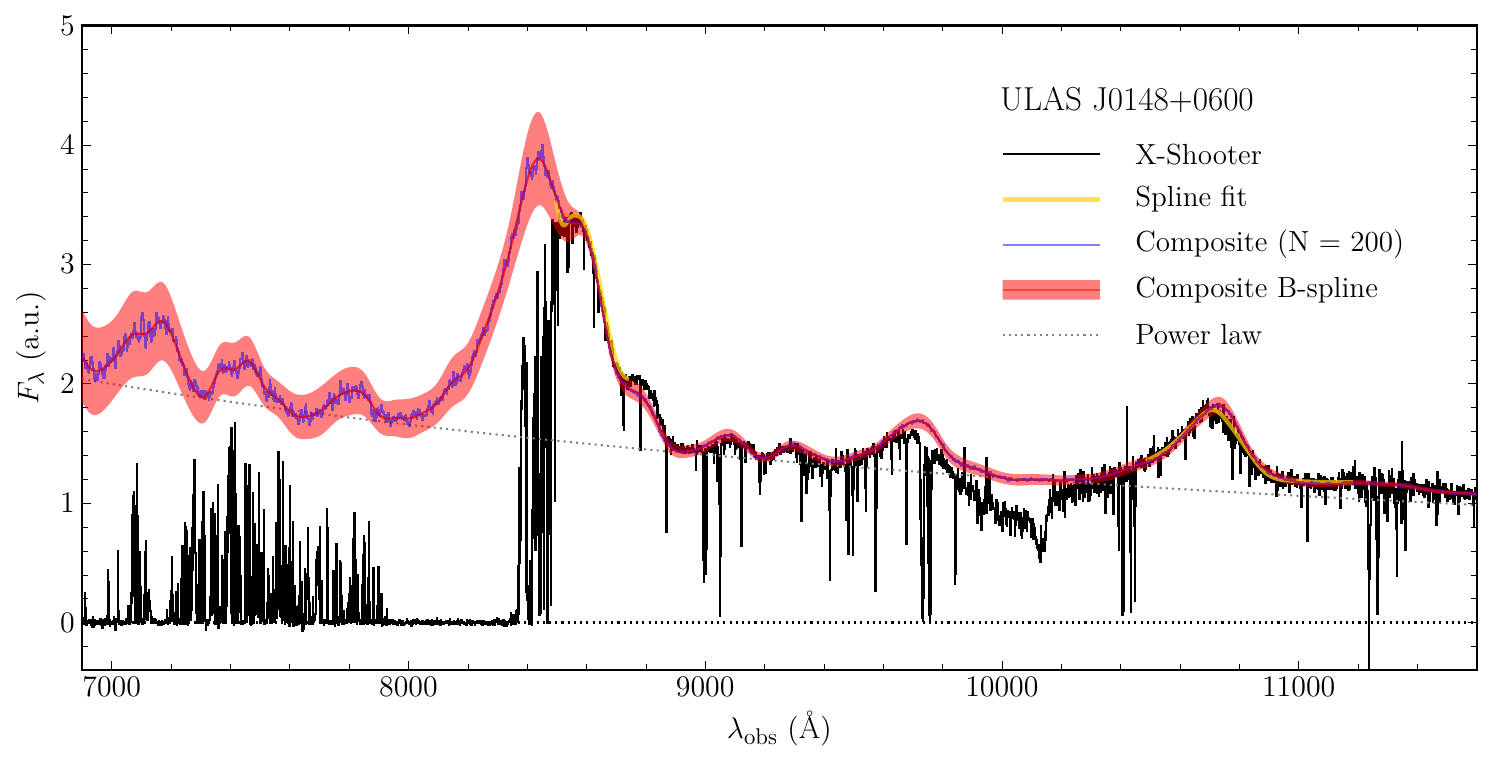}
   \vspace{-0.05in}
   \caption{X-Shooter spectrum of ULAS J0148$+$0600.  The black line shows the observed flux, in arbitrary $F_\lambda$ units, binned to 50 \kms\ pixels for display.  The short yellow lines are spline fits to the flux over the red side of the \lya+\nv\ emission line and the blueshifted \civ\ line.  The blue line is a composite continuum constructed from the spectra of 200 lower-redshift quasars chosen to match the spline fits.  The continuous red line and shaded region show a B-spline fit to the composite and its 1$\sigma$ uncertainty.  See text for details.  The dotted line shows our adopted power-law continuum for the non-ionizing UV emission, $L_{\nu} \propto \nu^{-0.6}$ ($F_{\lambda} \propto \lambda^{-1.4}$).  The depression in the flux with respect to the composite near $10000$~\AA\ is \civ\ broad absorption.  The 110 $h^{-1}$ comoving Mpc \lya\ trough appears as the region of zero flux over 7930 to 8362~\AA.  The transmission feature examined in this work is centered at $\sim$8385~\AA.}
   \label{fig:J0148_spec}
   \end{center}
   \end{minipage}
\end{figure*}

In this work, and in a companion paper by Zhu et al. (in prep), we present the first direct evidence that at least some of the extended \lya+\lyb\ troughs at $z < 6$ do indeed arise from neutral islands.  Both works demonstrate that these troughs are associated with damping wing absorption, the hallmark of neutral gas.  Zhu et al. use stacks to show statistical evidence of damping wing absorption in the  \lya\ forest adjacent to dark gaps.  Here we show that one specific dark gap produces a damping wing, and argue that the damping absorption is more likely to be caused by an extended neutral island than by a compact absorber such as a damped \lya\ system (DLA).

The dark gap we examine is part of the giant \lya\ trough towards ULAS~J0148$+$0600 (herein J0148).  This trough is remarkable partly because it is the longest ($\Delta \ell \simeq 1110$~\hinvMpc) and most opaque ($\tau_{\rm eff} \gtrsim 7$)\footnote{$\tau_{\rm eff} = -\log{F}$, where $F$ is the continuum-normalized transmission.} absorption feature known at $z < 6$ \citep{becker2018}.  As we describe below, it is also remarkable because the red end of the trough is adjacent to an unusually strong and smooth transmission feature at the end of the quasar proximity zone.  We will show that the characteristics of this feature strongly indicate the presence of damping wing absorption.  

This work shares similarities with studies that have identified damping wings in the spectra of quasars at $z > 7$ \citep{mortlock2011,greig2017,greig2019,greig2022,banados2018,davies2018,wang2020,yang2020}, and recent efforts that have extended this search down to $z = 6$ \citep{durovcikova2024}, but there are two key differences.  First, the damping wing signature is at $z < 6$, where no direct detection of neutral gas in the IGM has yet been made.  Second, the evidence for a damping wing comes from \lya\ transmission in the proximity zone, rather than from the otherwise unabsorbed continuum redward of the \lya\ forest (although \citet{davies2018} include both the proximity zone and the red-side continuum).  \citet{malloy2015} explored stacking $z < 6$ dark gaps to statistically detect damping wings in the presence of strong \lya\ forest absorption.  This is the approach taken by Zhu et al. (in prep), while here we focus on the damping wing from a single \lya\ trough.  Our work is perhaps most closely related to \citet{mesinger2004,mesinger2007}, who argued that the distribution of \lya\ fluxes near the proximity zone boundaries in three $z > 6.2$ quasars favors cosmological \hii\ regions within a significantly neutral IGM.  Here we go a step further, showing that the \lya\ transmission at the end of the J0148 proximity zone is consistent with damping wing absorption, while the no-damping-wing case is strongly ruled out.

The rest of the paper is organized as follows.  In Section~\ref{sec:J0148} we describe the J0148 transmission feature of interest and introduce the statistic we use to test for the presence of a damping wing.  We generate mock quasar proximity zones in Section~\ref{sec:mocks}, and show that generating features similar to the one towards J0148 requires the proximity zone transmission to be modified by damping wing absorption.  In Section~\ref{sec:discussion} we discuss the physical origin of the damping wing, arguing that it is more likely to arise from an extended neutral island rather than a compact absorber.  We also describe a scenario in which the transmission window signals the  recent passage of an intergalactic ionization front.  Finally, we summarize our results in Section~\ref{sec:summary}.  Throughout we assume a $\Lambda$CDM cosmology with $(\Omega_{\rm m}, \Omega_\Lambda, \Omega_{\rm b}, h) = (0.308, 0.692, 0.0482, 0.678)$ \citep{planckcollaborationxvi2014}.  Distances are quoted in units of comoving $h^{-1}$ Mpc, unless otherwise noted.

\section{The ULAS J0148$+$0600 \lya\ transmission window}\label{sec:J0148}

\begin{figure*}
   \centering
   \begin{minipage}{\textwidth}
   \begin{center}
   \includegraphics[width=1.0\textwidth]{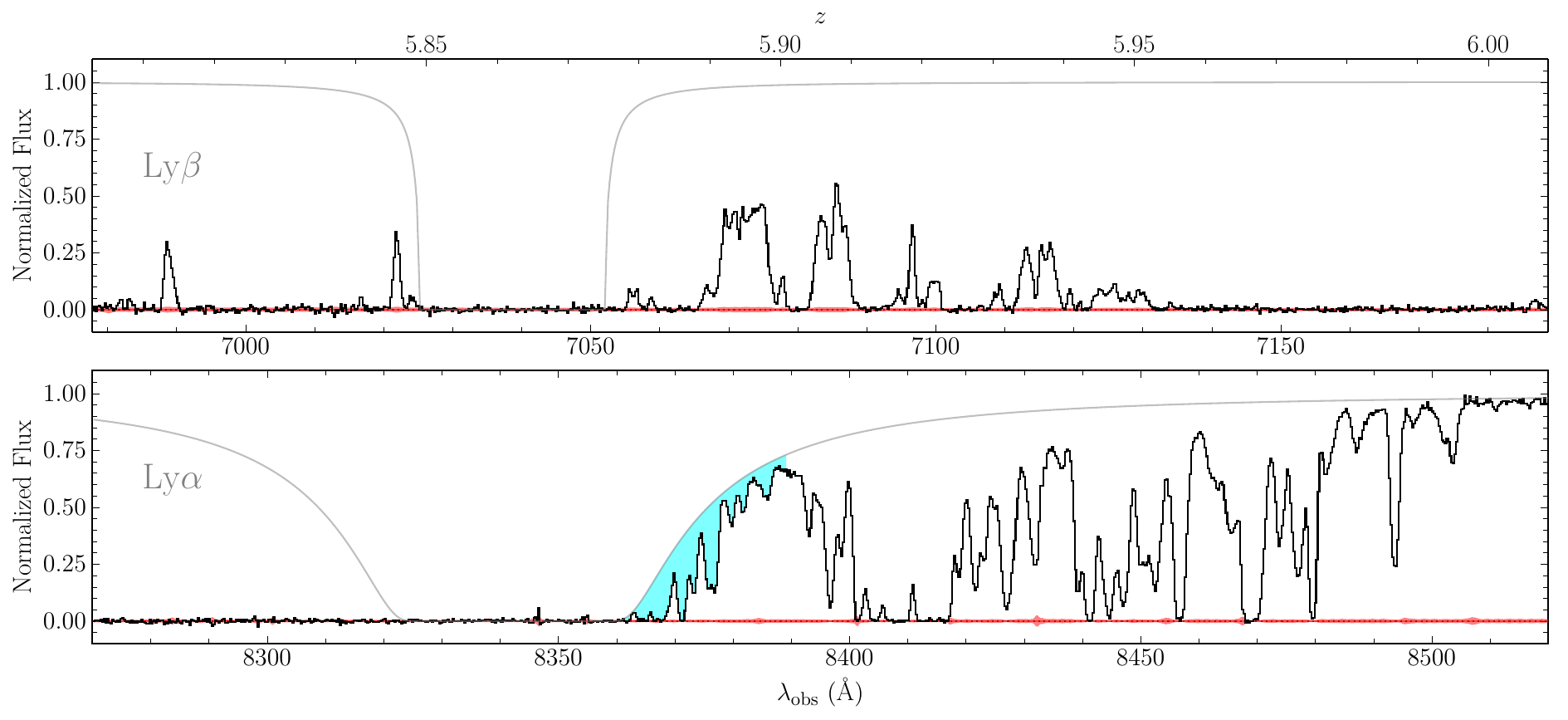}
   \vspace{-0.05in}
   \caption{J0148 proximity zone.  The top and bottom panels show the regions of the \lyb\ and \lya\ forests, respectively, corresponding to the proximity zone, where the wavelengths have been scaled to align the two panels vertically in redshift (top axis of \lyb\ panel).  The continuum-normalized flux is plotted as a histogram, with the 1$\sigma$ uncertainty in red shading.  The smooth line in both panels shows the damping profile for a neutral island of mean density with a length of 7.5 \hinvMpc, the maximum extent allowed by the bordering \lya\ and \lyb\ transmission.  The cyan shaded region in the lower panel shows the difference between the damping profile and the flux over a 1000~\kms\ region at the red edge of the damping wing.}
   \label{fig:J0148_prox}
   \end{center}
   \end{minipage}
\end{figure*}

The primary feature examined in this paper is an extended window of \lya\ transmission centered at $\lambda \simeq 8385$~\AA, which falls at the blue end of the J0148 proximity zone.  Here we describe the J0148 spectrum used in this work, introduce our technique for  continuum normalization, and describe the basic characteristics of the transmission window.  We also introduce a statistic, \meandiff, that will be used to test for the presence of damping wing absorption.

Our analysis is based on a 10-hour VLT X-Shooter spectrum of J0148 first presented in \citet{becker2015} and more recently included in the extended XQR-30 sample of \citet{dodorico2023}.  We refer the reader to those paper for details of the data reduction, but note that the J0148 spectrum is extremely deep, with a continuum-to-noise ratio (based on the continuum described below) of $\sim$130--220 per 10~\kms\ pixel over the transmission window and the adjacent \lya\ trough.  The spectrum is shown in Figure~\ref{fig:J0148_spec}.  We adopt a redshift for J0148 of $z_{\rm q} = 5.9896$, which is based on [\ion{C}{ii}] 158 $\mu$m emission-line measurements (Boseman et al., in prep).  

We normalize the spectrum using a continuum that is based on a composite of lower-redshift objects.  The method is described in detail in Appendix~\ref{sec:continuum}.  Briefly, we identify quasars from the Sloan Digital Sky Survey Data Release 16 \citep[DR16;][]{lyke2020} that are good matches to J0148 in terms of their transmission in two wavelength intervals redward of the \lya\ forest.  The intervals cover the red side of the \lya+\nv\ emission line and the \civ\ emission line, which are highlighted yellow in Figure~\ref{fig:J0148_spec}. 
\lya\ emission is of particular importance for this work because the transmission window of interest lies on the blue side of the \lya\ line.  We also include \civ\ in the selection because the strength and systemic velocity offset of \civ\ emission are known to be correlated with those of \lya\ \citep[e.g.,][]{richards2011,greig2017,davies2018}.  
We select the 200 best-matching DR16 spectra based on mean squared residuals with J0148 within these intervals.  Each spectrum is then corrected for foreground \lya\ and \lyb\ absorption using a statistical approach that includes reduced absorption within the proximity zone (see Appendix~\ref{sec:continuum} for details).  A composite is computed by averaging the corrected spectra as a function of their rest-frame wavelengths (blue histogram in Figure~\ref{fig:J0148_spec}).  Our final continuum model is a b-spline fit to this composite (red solid line), with an uncertainty (red shaded region) that is calculated from the scatter in the individual corrected spectra, smoothed on 5000~\kms\ scales.  At the wavelength of the transmission window we find a 1$\sigma$ continuum uncertainty of 10\%.  The uncertainty estimate is conservative in that part of this scatter will be due to variations in the amount of \lya\ forest absorption between lines of sight rather than intrinsic variations in the DR16 spectra.

The continuum-normalized J0148 spectrum covering the \lya\ proximity zone and the red end of the giant \lya\ trough is displayed in Figure~\ref{fig:J0148_prox}.  We also show the corresponding section of the \lyb\ forest.  The most conspicuous feature of the \lya\ proximity zone is a strong (maximum transmission $\simeq$ 0.7) and extended transmission window spanning $\lambda \simeq 8370$--8400~AA\ ($z \simeq 5.89$--5.91).  Ignoring peculiar velocities, the wavelength range corresponds to a line-of-sight extent of $\sim$7~\hinvMpc.  In addition to its amplitude and width, two features of the transmission window stand out.  First, it lies very near to the red end of the \lya\ trough, which is also dark in \lyb\ over $z = 5.849$--5.878.  Second, the transmission on the blue side increases relatively smoothly from zero up to its maximum.  

Motivated by the shape of this transmission window and the potential association of dark \lya+\lyb\ troughs with neutral islands, we calculate the transmission profile for a neutral island that could reside near the end of the J0148 proximity zone.  We assume that the island is at mean density, and that it spans the maximum extent allowed by the first \lya+\lyb\ dark gap blueward of the proximity zone.  The \lya\ and \lyb\ transmission at the red and blue ends of this gap, respectively, allow a 7.5~\hinvMpc\ island with its near edge roughly 34~\hinvMpc\ from the quasar.  No further constraints are imposed by higher-order Lyman-series transitions.  For reference, the central redshift of such an island is 5.862, and the total \hi\ column density is $\log{(N_{\rm H\, I} / {\rm cm^{-2}})} \simeq 20.5$.  We note, however, that the transmission  profile will differ from that of a compact absorber of the same column density due to the extended nature of the island \citep{miralda-escude1998}.  The \lya\ and \lyb\ transmission profiles for the potential island are shown in Figure~\ref{fig:J0148_prox}.  

The damping wing of the neutral island roughly traces the blue side of the broad \lya\ transmission window.  We also note that the entire \lya\ (and \lyb) proximity zone transmission falls under the damping profile.  The J0148 proximity zone is therefore consistent with the presence of a damping wing created by an extended neutral island.  The central question we will examine in this paper is whether a damping wing is actually required  to produce the transmission profile of the broad \lya\ window, or if such a feature can arise by chance from resonant absorption alone.

To address this question we will compare the J0148 proximity zone to mock quasar proximity zones generated with and without damping wing absorption.  The spectra will be evaluated based on their similarity to a damping wing profile at the blue end of the proximity zone.  Specifically, we will measure the mean difference, \meandiff, between the proximity zone transmission and the damping wing template for a neutral island similar to the one shown in Figure~\ref{fig:J0148_prox}.  In order to focus on the part of the spectrum most impacted by the damping wing, and to simplify the modeling of the proximity zones, we measure \meandiff\ over a 1000~\kms\ region that starts where the red side of the damping wing template first exceeds 0.01.  The difference between the J0148 transmission and the damping wing over this region is shaded in Figure~\ref{fig:J0148_prox}.  Without any adjustment of the profile parameters we find $\langle \Delta F \rangle = 0.15$ for J0148.  Below we test whether this value indicates that a genuine damping wing is likely to be present.

\section{Simulated proximity zones}\label{sec:mocks}

\subsection{Generating mock proximity zones}\label{sec:generating}

\begin{figure}
   \begin{center}
   \includegraphics[width=0.46\textwidth]{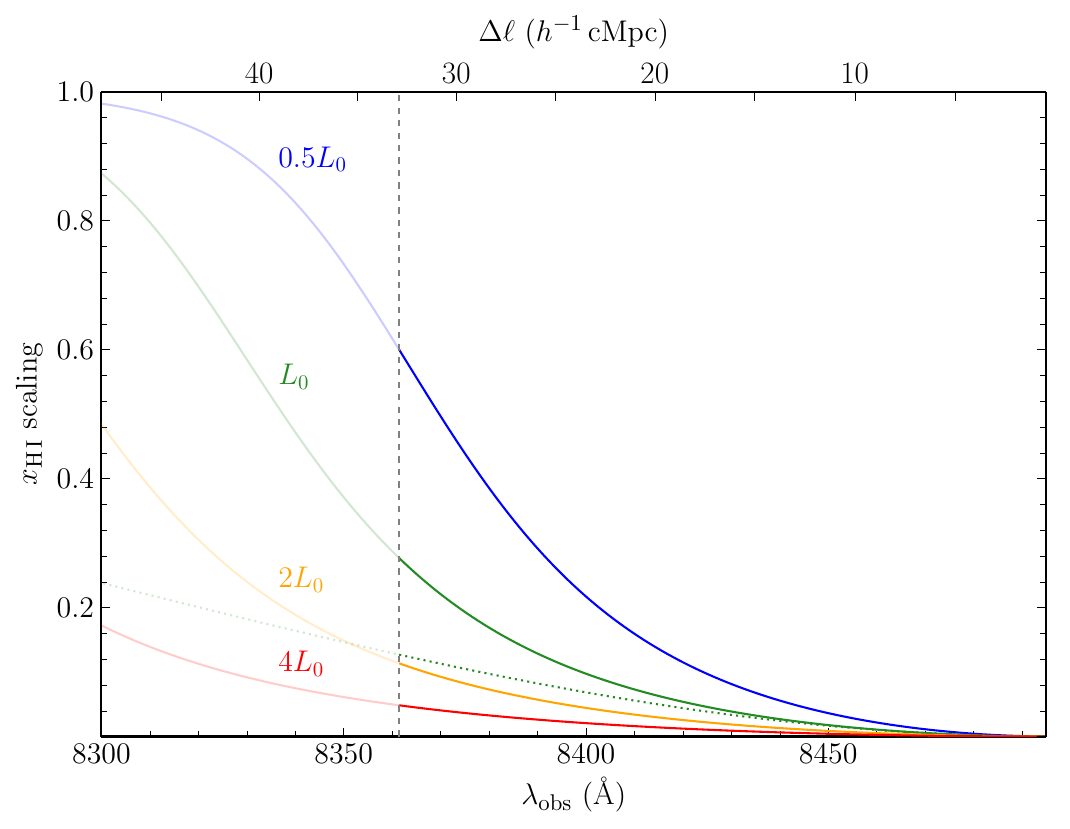}
   \vspace{-0.05in}
   \caption{Scaling of the \hi\ neutral fraction in the proximity zone models used in this work.  Distances are shown along the top axis, with the bottom axis showing the corresponding observed wavelength for a quasar at $z_{\rm q} = 5.9896$.  The solid lines show the factors by which the neutral fraction is multiplied as a function of distance from the quasar for four different quasar luminosities.  $L_0$, corresponds to the nominal ionizing UV luminosity of J0148.  The neutral fraction scaling is calculated for 0.5, 1.0, 2.0, and 4.0 times this luminosity (see text for details).  The vertical dashed line corresponds to the edge of transmitted \lya\ flux in the J0148 proximity zone. Note that these scalings are only applied to the ionized gas within the proximity zone;  trends out to larger distances are shown for illustration.  The scalings include the effect of ionizing opacity within the proximity zone, as discussed in Section~\ref{sec:generating}.  For reference, the dotted line shows the zero-opacity case for $L_0$.}
   \label{fig:prox_scaling}
   \end{center}
\end{figure}

Mock proximity zones are generated using one-dimensional lines of sight drawn from the $z = 6.0$ output of the Sherwood hydrodynamical simulation suite \citep{bolton2017}.  We use 5,000 lines of sight from the 80-2048 run, which has a box size of 80 \hinvMpc\ and 2048$^3$ gas particles with a mass of $7.97 \times 10^5~h^{-1}~{\rm M_\odot}$.  The lines of sight are not drawn through the massive halos expected to host bright quasars; however, this is unlikely to affect our analysis, which is focused on features that are offset by more than $\sim$30 \hinvMpc\ from the quasar position.  By comparison, the virial radius for a $10^{12}~h^{-1}~{M_\odot}$ halo will be $\sim$0.3~\hinvMpc\  at $z = 6$ \citep{barkana2001}.  We rescale the gas temperatures using a power-law temperature-density relation of the form $T(\rho) = T_0 (\rho / \langle \rho \rangle)^{\gamma-1}$, adopting $T_0 = 12000$~K and $\gamma = 1.2$, broadly consistent with measurements from \citet{gaikwad2020} over $5.3 < z < 5.9$.  Baseline \lya\ optical depths are set by globally rescaling the hydrogen neutral fractions to produce a mean \lya\ transmission of 0.01, similar to the measurements of \citet{bosman2021} at $z = 5.9$ (i.e., near the end of the J0148 proximity zone).  For each of the samples described below we then generate two proximity zones per line of sight, one in each direction, for a total of 10,000 proximity zones per sample.    

The simulated proximity zones consist of an extended optically thin photo-ionized region that terminates at an optically thick absorber, with various absorbers considered in the following sections.  For the optically thin portions we use the proximity zone model described in \citet{becker2021} and \citet{zhu2023}.  Briefly, hydrogen neutral fractions along the line of sight are scaled according to $x_{\rm H\, I} \propto (\Gamma / \Gamma_{\rm bg})^{-1}$, where the local photoionization rate, $\Gamma = \Gamma_{\rm bg} + \Gamma_{\rm q}(r)$, combines contributions from the diffuse UV background and the ionizing flux from the quasar.  The radial dependence of the quasar flux includes both a $1 / r^2$ geometric dilution and attenuation due to the ionizing opacity within the proximity zone.  The opacity scales with the local photoionization rate as $\kappa_{912} = \kappa_{912}^{\rm bg} ( \Gamma / \Gamma_{\rm bg} )^{\xi}$, and therefore increases with distance from the quasar following equation (8) in \citet{becker2021}.  We adopt a background opacity value of $\log{(\kappa_{\rm bg} / ({\rm cm}^{-1})} = -24.5$.  This corresponds to a mean free path of ionizing photons of 1.0 proper Mpc, consistent with recent measurements from \citet{becker2021} and \citet{zhu2023}.  For $\xi$ we adopt 0.67 based theoretical models of the ionizing sinks \citep[][see discussions in \citealt{davies2016,daloisio2018,daloisio2020,becker2021}]{furlanetto2005,mcquinn2011}.  For the background photoionization rate we adopt $\Gamma_{\rm bg} = 0.17 \times 10^{-12}~{\rm s^{-1}}$, consistent with constraints from \citet{gaikwad2023} at $z = 5.9$, albeit slightly higher than their nominal value.  For the simulated quasars we adopt a nominal absolute UV magnitude at rest-frame 1450~\AA\ of $M_{1450} = -27.4$, similar to J0148 \citep{banados2016}.  The UV spectrum is modeled as a double power law of the form $L_\nu \propto \nu^{-\alpha}$, with $\alpha = 0.6$ betwen 1450 \AA\ and 912 \AA, and 1.5 shortward of 912 \AA\ \citep[see][and references therein]{lusso2015,becker2021}.  This agrees well with our composite-based continuum over the non-ionizing UV part of the spectrum (Figure~\ref{fig:J0148_spec}).

In principle there are a variety of factors that will impact the neutral fraction within the proximity zone.  These include the full quasar spectral energy distribution, the quasar lifetime and/or variability, variations in ionizing opacity along the line of sight not captured by the model above, fluctuations in the ionizing background, and gas temperature.  The neutral fraction at the edge of the proximity zone will also depend on the distance from the quasar.  In practice, however, we will focus only on a narrow region at the end of the proximity zone, for which the important quantity is the local neutral fraction.  We therefore sample a range of neutral fractions by simply varying an ``effective'' quasar luminosity, $L_{\rm q}$, but note that these variations may also result from other factors, which we examine further in Section~\ref{sec:charachteristics}.  The scaling of $x_{\rm H\,I}$ for optically thin regions within the proximity zones is shown in Figure~\ref{fig:prox_scaling}.  We refer to the scaling given by the nominal parameters above as $L_{\rm q} = L_0$, and compute proximity zones for $L_{\rm q} = 0.5 L_0$, $L_0$, 2.0$L_0$, and 4.0$L_0$.  At the distance to the edge of the J0148 proximity zone,  $\sim$33 \hinvMpc, the neutral fraction scaling ranges from 0.60 (i.e., 60\% of the value in the absence of the proximity effect) for $0.5 L_0$, to 0.05 for $4 L_0$.  As shown below, this range is sufficient to evaluate the role of ionization in the proximity zone models we consider.   We note that the scaling of the neutral fraction with quasar luminosity is amplified by the impact of $L_{\rm q}$ on the ionizing opacity, following the discussion above.  The total optical depth within the proximity zone is $\tau \simeq 1.66$, 0.98, 0.57, and 0.34 for $L_{\rm q} = 0.5L_0$, $L_0$, $2L_0$, and $4L_0$, respectively.  For reference, we plot the zero-opacity case for $L_0$ as a dotted line in Figure~\ref{fig:prox_scaling}.

Finally, we apply a random continuum error to each simulated line of sight.  The continuum error is modeled as a constant factor drawn from a normal distribution with a mean of 1.0 and a standard deviation of 0.1, similar to the fractional continuum uncertainty of J0148 over the transmission window at $\sim$8385~\AA\ (Figure~\ref{fig:J0148_prox}).  We note that this does not include a possible wavelength dependence in the continuum; however, the analysis below focuses on a narrow wavelength range, for which a uniform continuum rescaling should be adequate.  We convolve each spectrum with a Gaussian kernel with full width at half maximum of 23~\kms, equal to the measured X-Shooter resolution for J0148 \citep{dodorico2023}.  We also rebin the spectra to a pixel scale similar to the J0148 spectrum, although the results should not depend sensitively on the choice of binning. 

\subsection{\meandiff\ measurement method}

We evaluate the proximity models below based on how frequently they produce \meandiff\ values equal to or less than the J0148 value.  As with our analysis of J0148, we use a damping profile template calculated from a 7.5 \hinvMpc\ neutral island at mean density.  \meandiff\ is calculated as the difference between the proximity zone transmission and this template over a 1000~\kms\ region starting where the red wing of the damping profile first exceeds 1\%.  Conservatively, we minimize \meandiff\ for each line of sight by placing the island at the minimum distance from the quasar where the transmission within this region does not exceed the damping profile.  

We note that we allow the proximity zone transmission to exceed the damping profile at  wavelengths redder than the \meandiff\ region, for multiple reasons.  First, there may be wavelength-dependent errors in the continuum that are not captured by our approach of rescaling the transmission in each simulated proximity zone by a constant.  Second, there may be variations in the $x_{\rm H\,I}$ scaling that are not well modeled by the radial profiles plotted in Figure~\ref{fig:prox_scaling}.  Third, the simulated lines of sight are drawn from random locations with the simulation box and do not capture the enhanced densities that may be present in the vicinity of a bright quasar, although this would mainly impact absorption close to the quasar redshift (see Section~\ref{sec:generating}).  Restricting our analysis to a narrow region at the edge of the proximity allows us to simplify our analysis by avoiding these more nuanced aspects of the modeling.  In the future, however, a more robust modeling of the transmission and continuum could allow the full proximity zone to be used to constrain the presence of a damping wing \citep[e.g.,][]{davies2018}.  We note here that for our nominal continuum estimate, the J0148 flux remains below the damping wing profile over the entire proximity zone (Figure~\ref{fig:J0148_prox}).

\begin{figure}
   \begin{center}
   \includegraphics[width=0.46\textwidth]{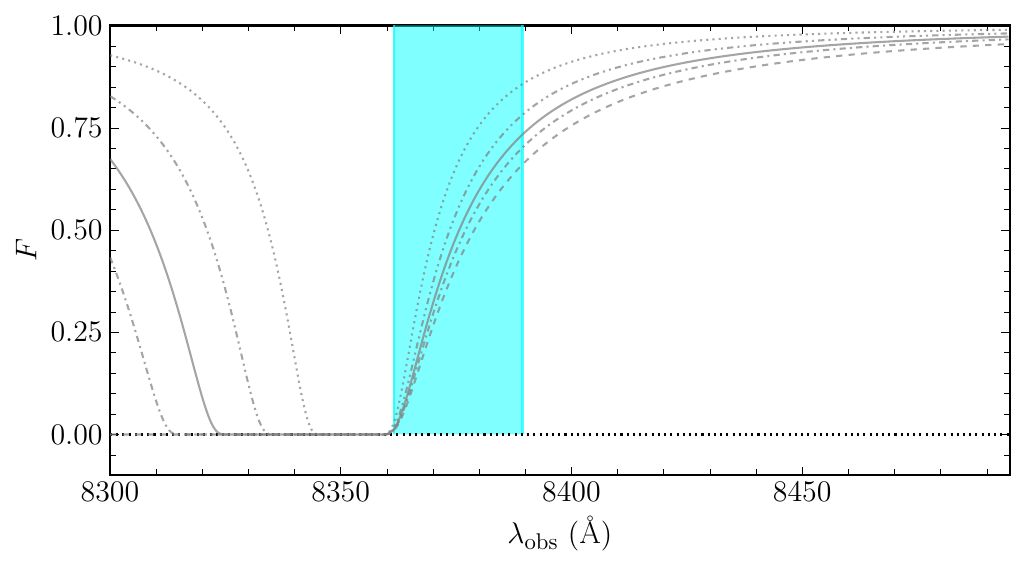}
   \vspace{-0.05in}
   \caption{Transmission profiles for mean-density neutral islands of length 2.5 (dot), 5.0 (dash-double-dot), 7.5 (solid), 10 (dash-dot) and 15 (dash) \hinvMpc.  The near edge for all islands is 34~\hinvMpc\ from a quasar at $z_{\rm q} = 5.9896$.  The shaded region shows the 1000 \kms\ interval used to calculate \meandiff.  We use the 7.5~\hinvMpc\ template for our primary analysis, which is motivated by the \lya\ and \lyb\ transmission near the end of the J0148 proximity zone.  Our results do not substantially change, however, when other templates are used.}
   \label{fig:profiles}
   \end{center}
\end{figure}

It is worth considering whether the details of our \meandiff\ measurement may impact the results below.  For example, we have chosen to measure \meandiff\ over a 1000~\kms\ region partly because this is the extent to which the J0148 transmission window appears to follow the \lya\ damping wing in Figure~\ref{fig:J0148_prox}.  Our main result, that we exclude models without damping wings (see Section~\ref{sec:LLS} below), actually becomes somewhat stronger if we measure \meandiff\ out to 1200~\kms, and only mildly weaker at 1500 \kms, which roughly corresponds to the red edge of the J0148 transmission window.  Predictably, \meandiff\ becomes less able to discriminate between models if we extend the region further to 2000~\kms\ due to the contribution from strong absorption over 8401--8417~\AA.  It is expected that resonant \lya\ forest absorption will become increasingly dominant over damping wing absorption at large velocities, making \meandiff\ a less sensitive statistic over wider windows.  For this line of sight, therefore, we have chosen 1000~\kms\ in order to provide a reasonable baseline to test for a damping wing signal while limiting the contamination from unrelated absorption.

We also note that the results below do not depend sensitively on the damping wing template used to measure \meandiff.  As described above, our choice of a template generated from a 7.5~\hinvMpc\ neutral region at mean density is motivated by the spacing between \lya\ and \lyb\ transmission peaks in the J0148 spectrum.  For comparison, in Figure~\ref{fig:profiles} we show damping profiles for mean-density neutral islands with lengths ranging from 2.5~\hinvMpc\ to 15~\hinvMpc, which are aligned at their edges nearest to the quasar.  We note that contiguous islands longer than $\sim$7.5~\hinvMpc\ are not allowed towards J0148 by the \lyb\ transmission peaks in Figure~\ref{fig:J0148_prox}; however, there could be multiple islands separated by small gaps.  The extent of the absorption trough depends strongly on the size of the island, but \meandiff\ depends only on the shape of the red wing, for which the variation is more modest.  We repeated the tests below using these other profiles and found that the main results do not change substantially.   We also tried marginalizing over the template island length by taking the minimum \meandiff\ from among the profiles shown in Figure~\ref{fig:profiles} and again found no substantial change.

\subsection{No damping wing absorption}\label{sec:LLS}

\begin{figure*}
   \centering
   \begin{minipage}{\textwidth}
   \begin{center}
   \includegraphics[width=0.95\textwidth]{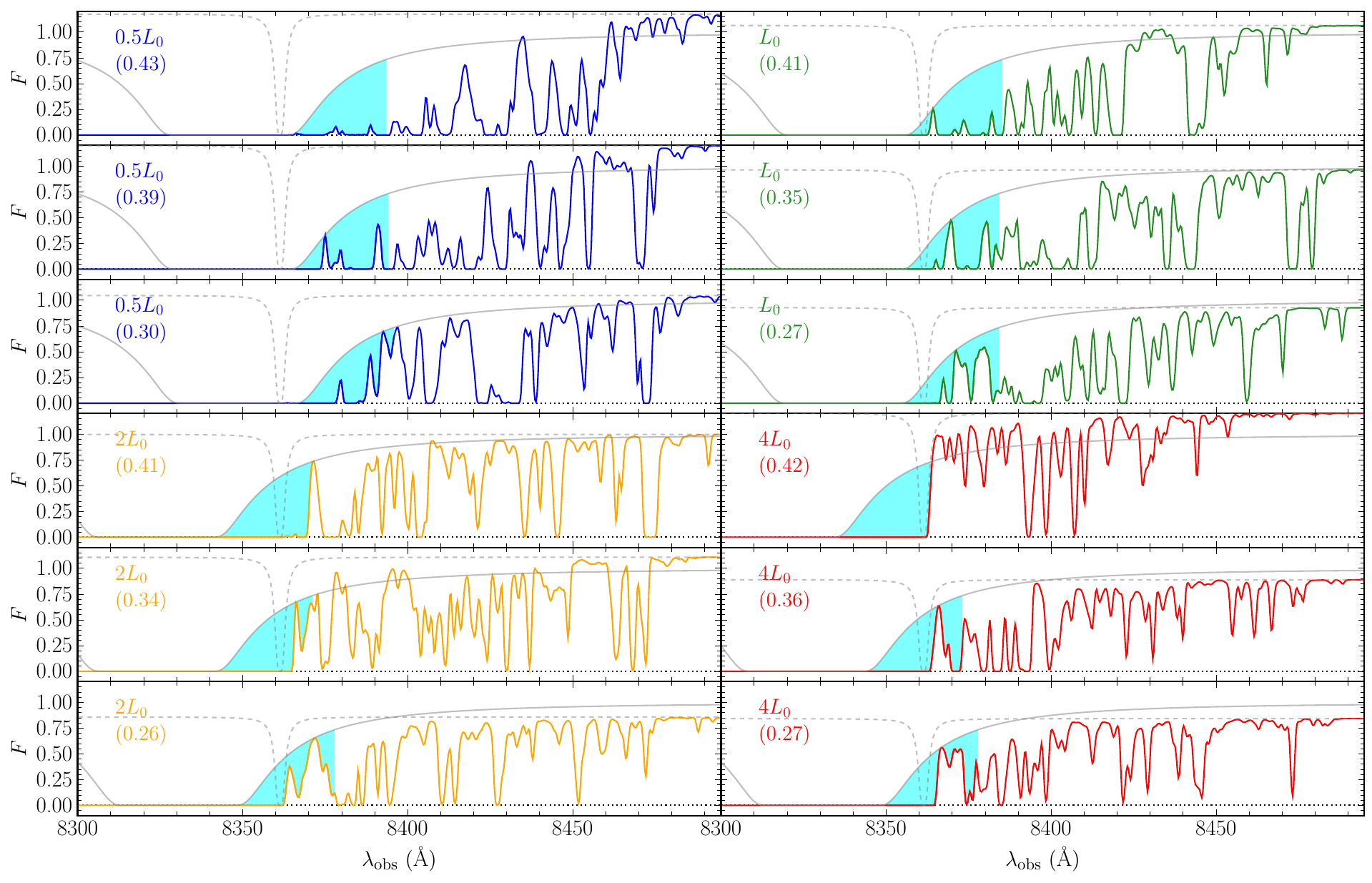}
   \vspace{-0.05in}
   \caption{Examples of simulated proximity zones ending at Lyman limit systems with column density $\log{(N_{\rm H\, I} / {\rm cm^{-2}})} = 18.0$.  The transmitted flux in each panel has been rescaled by a random amount to emulate continuum errors (see text for details).  Dashed lines show the LLS absorption profile, which has been rescaled by the same amount. The smooth solid line shows the absorption profile for a 7.5~\hinvMpc\ neutral island at mean density.  The shaded region shows the difference between this profile and the transmitted flux over a 1000~\kms\ region.  For each line of sight, the damping profile is shifted in wavelength such that it minimizes \meandiff\ while not allowing the transmission to exceed the damping profile within the widow.  The transmission is allowed to exceed the damping profile at longer wavelengths, however (see text for details).  There are three lines of sight plotted for each quasar luminosity scaling in Figure~\ref{fig:prox_scaling}, where the scaling is indicated in the upper left corner of each panel.  The lines of sight correspond to the 5$^{\rm th}$, 50$^{\rm th}$, and 95$^{\rm th}$ percentiles (top to bottom) in the \meandiff\ distribution for each $L_{\rm q}$.  The value of \meandiff\ is given in parentheses.}
   \label{fig:LLS_los}
   \end{center}
   \end{minipage}
\end{figure*}

\begin{figure}
   \begin{center}
   \includegraphics[width=0.43\textwidth]{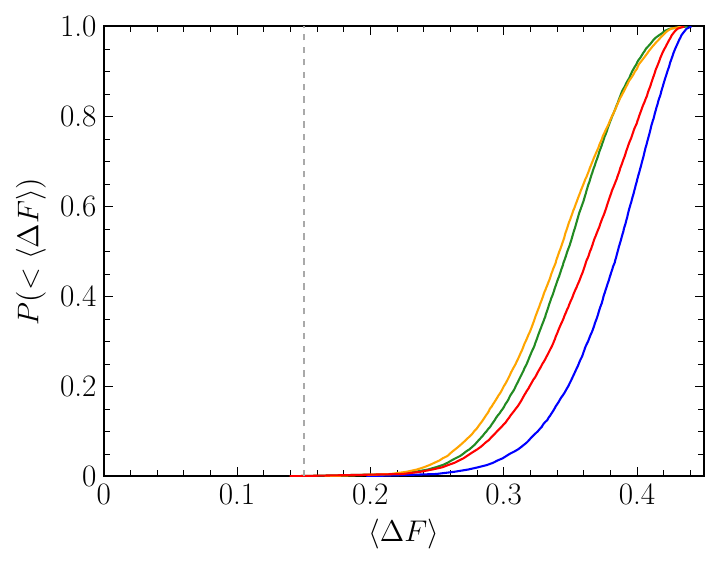}
   \vspace{-0.05in}
   \caption{Cumulative probability distributions of \meandiff\ for simulated lines of sight ending in Lyman limit systems.  The solid lines show $P(< \langle \Delta F \rangle)$ for the four quasar luminosity scalings, with colors correspond to Figure~\ref{fig:prox_scaling}.  For reference, the left-to-right ordering of these lines at $P(< \langle \Delta F \rangle) = 0.5$ is $L_{\rm q} = 2L_0, L_0, 4L_0, 0.5L_0$.  The vertical dashed line shows \meandiff\ for J0148.}
   \label{fig:LLS_stats}
   \end{center}
\end{figure}

We first test whether features similar to the J0148 transmission window can occur in the absence of damping wing absorption.  For this test, the simulated proximity zones are truncated at a Lyman limit system (LLS) with neutral hydrogen column density $\log{(N_{\rm H\, I} / {\rm cm^{-2}})} = 18.0$.  This provides sufficient optical depth to truncate the quasar's ionizing flux but will not produce a significant damping wing.  We assume that the LLS should reside in a density peak that is at least three times the mean density, but find that either increasing this to ten times the mean or removing the density requirement entirely would have little impact on our results.  For each line of sight we randomly select a density peak, and then shift the periodic line of sight such that this peak lies at $z = 5.878$ when peculiar velocities are taken into account, assuming the quasar is at $z = 5.9896$.  The LLS redshift is chosen to match the maximum extent of the J0148 proximity zone transmission, and for zero peculiar velocity corresponds to a distance from the quasar of 33 \hinvMpc.  We set $\Gamma_{\rm q} = 0$ beyond the LLS.  We also produce a complete absorption trough outside of the proximity zone by arbitrarily increasing  $x_{\rm H\, I}$.  For this test we remain agnostic about the physical origin of the trough, although one may arise from an ionized IGM if the UV background outside of the proximity zone is highly suppressed.  Here a trough is introduced simply to maximize the similarity of a simulated line of sight to J0148.  

We note that a few lines of sight randomly intersect high column density absorbers that are native to the simulation box.  These absorbers naturally produce damping wing absorption, which can impact the appearance of the proximity zone.  Because we wish to determine whether a feature similar to J0148 can be produced in the absence of any damping wing absorption, we suppress the Lorentzian wings by using a Gaussian absorption profile for ionized gas (although not for the LLS).  The impact of damping wing absorption from higher-column density absorbers is examined below.

Example proximity zones ending in LLSs are plotted in Figure~\ref{fig:LLS_los}.  We plot three lines of sight for each \Lquasar\ scaling, which are chosen to lie at the 5$^{\rm th}$, 50$^{\rm th}$, and 95$^{\rm th}$ percentiles in \meandiff\ for that scaling (bottom to top for each \Lquasar).  Similar to Figure~\ref{fig:J0148_prox}, the template damping profile is plotted in grey and the shaded region shows the 1000 \kms\ region over which \meandiff\ is measured.  The full \meandiff\ distributions are plotted in Figure~\ref{fig:LLS_stats}.  The smallest \meandiff\ values tend to be produced by $L_{\rm q} = L_0$ and $2L_0$.  By comparison, $0.5L_0$ lines of sight tend to exhibit only weak transmission peaks at the edge of the proximity zone, while the strong transmission for $4L_0$ tends to prohibit the \meandiff\ region of the damping profile from extending over the transmission peaks.  In all cases, however, the the LLS models fail to generate \meandiff\ values equal to or less than the J0148 value in more than 0.01\% of cases.  We therefore conclude that lines of sight without significant damping wings are highly unlikely to produce the kind of smooth \lya\ transmission feature seen at the blue end of the J0148 proximity zone.

\subsection{Compact damped \lya\ absorbers}\label{sec:DLA}

\begin{figure*}
   \centering
   \begin{minipage}{\textwidth}
   \begin{center}
   \includegraphics[width=0.95\textwidth]{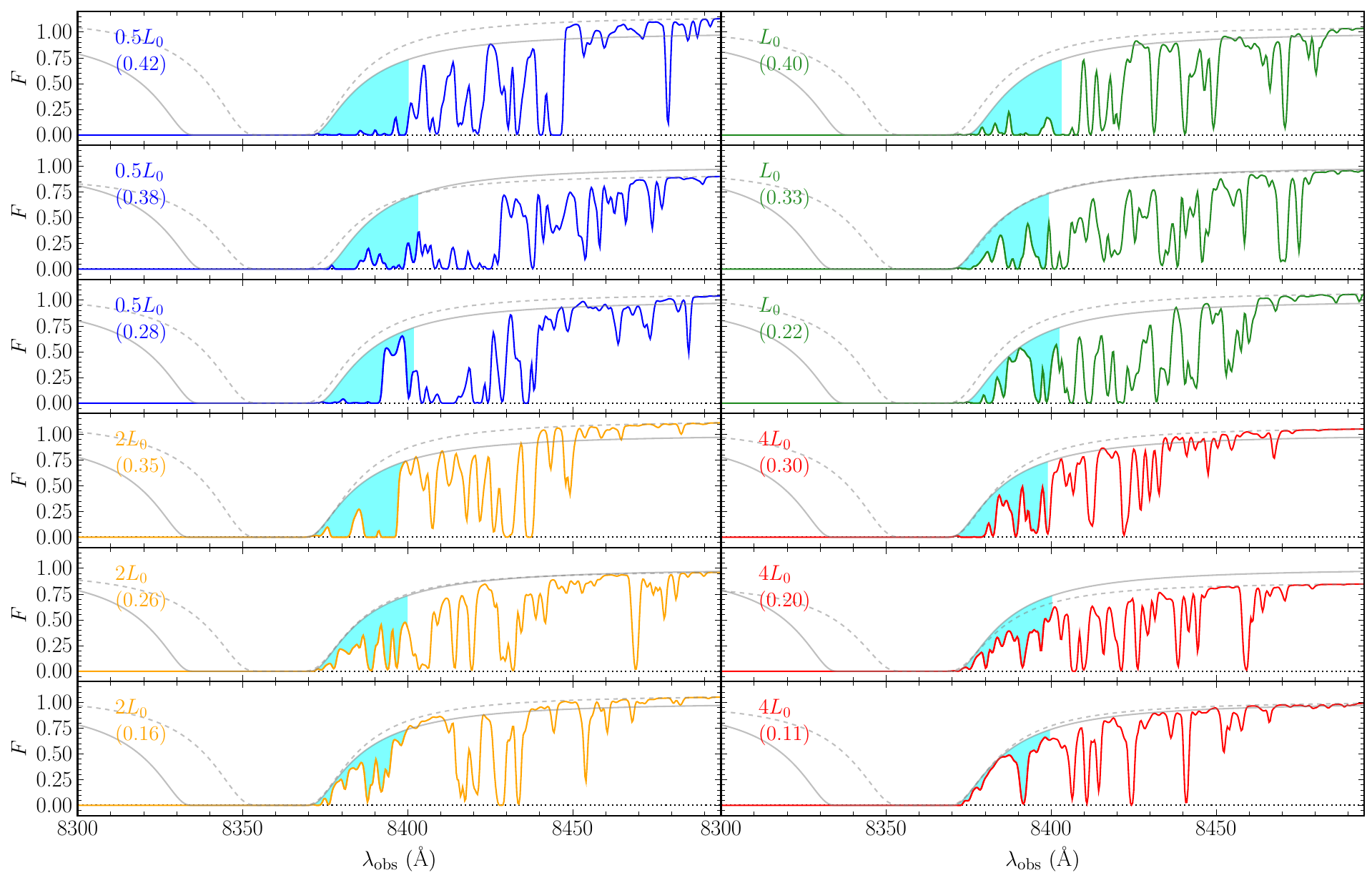}
   \vspace{-0.05in}
   \caption{Similar to Figure~\ref{fig:LLS_los}, but for simulated lines of sight ending in compact DLAs with column density $\log{(N_{\rm H\, I} / {\rm cm^{-2}})} = 20.3$.  The dashed line in each panel shows the DLA damping profile, which has been rescaled by the same amount as the transmitted flux.  The smooth solid line is a template profile for a  mean-density 7.5~\hinvMpc\ neutral island that has been shifted to produce the minimum allowable \meandiff\ over the 1000 \kms\ shaded region at the red edge of the template damping wing.}
   \label{fig:DLA_los}
   \end{center}
   \end{minipage}
\end{figure*}

\begin{figure}
   \begin{center}
   \includegraphics[width=0.43\textwidth]{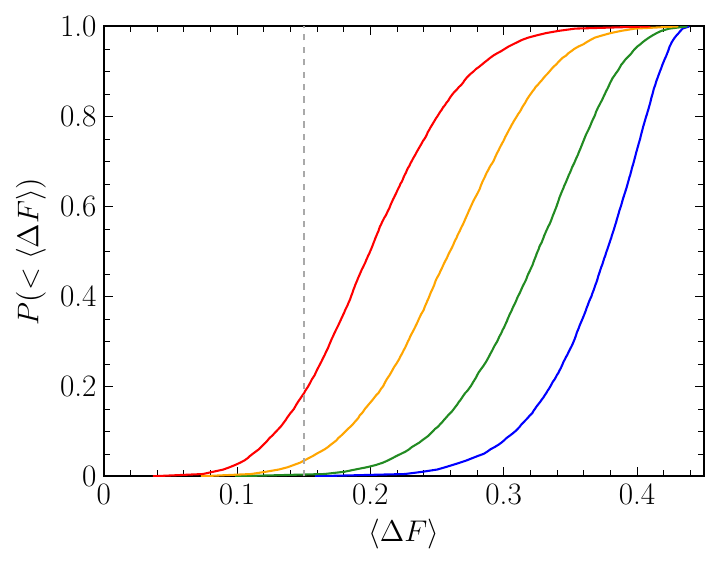}
   \vspace{-0.05in}
   \caption{Similar to Figure~\ref{fig:LLS_stats}, but for simulated proximity zones ending in compact DLAs with column density $\log{(N_{\rm H\, I} / {\rm cm^{-2}})} = 20.3$.   Colors correspond to the quasar luminosity scalings in Figure~\ref{fig:prox_scaling}.  The left-to-right ordering of the solid lines is $L_{\rm q} = 4L_0, 2L_0, L_0, 0.5L_0$.  The vertical dashed line shows \meandiff\ for J0148.} 
   \label{fig:DLA_stats}
   \end{center}
\end{figure}

Next we evaluate whether the J0148 feature could be created via damping wing absorption from a compact source such as a damped \lya\ system (DLA) residing in a galaxy halo.  Here we reuse the simulated lines of sight generated in Section~\ref{sec:LLS}, but add higher-column density absorbers to the end of the proximity zone ($z=5.878$).  The \hi\ column density is varied from $\log{(N_{\rm H\, I} / {\rm cm^{-2}})} = 19.0$ to 20.5 in increments of 0.1 dex, thus spanning a broad range of DLAs and sub-DLAs.  The minimum \meandiff\ value is then found for each combination of $N_{\rm H\, I}$ and \Lquasar.  In this section we focus on $\log{(N_{\rm H\, I} / {\rm cm^{-2}})} = 20.3$ because it is the canonical DLA value and close the column density where the probability of producing \meandiff\ values at or below that of J0148 tends to be greatest.  The full range of column densities is examined in Appendix~\ref{sec:compact}.

Example lines of sight ending in compact DLAs with column density $\log{(N_{\rm H\, I} / {\rm cm^{-2}})} = 20.3$ are shown in Figure~\ref{fig:DLA_los}.  Similar to Figure~\ref{fig:LLS_los}, lines of sight at the 5$^{\rm th}$, 50$^{\rm th}$, and 95$^{\rm th}$ percentiles in \meandiff\ are shown for each \Lquasar\ scaling.  The full \meandiff\ distributions are plotted in Figure~\ref{fig:DLA_stats}.  We find that a significant percentage of lines of sight with \meandiff\ values equal or less than the J0148 value are  produced for the higher \Lquasar\ values (3\% for 2\Lquasar, 18\% for 4\Lquasar).  As seen in Figure~\ref{fig:DLA_los}, in cases where \meandiff\ falls near or below the J0148 value of 0.15, the red wing of the neutral island damping wing template (solid grey line) tends to closely follow the red wing of the actual DLA profile (dashed line), even though the template profile is considerably more extended on the blue side.  In these cases, the DLA profile modifies a span of the proximity zone that otherwise has high transmission within the \meandiff\ region, as seen by the nearness of the tops of the transmission peaks to the DLA profile.  We note that these lines of sight also exhibit considerably more transmission over the full proximity zone than is seen in J0148 (e.g., comparing the bottom panel of Figure~\ref{fig:J0148_prox} to the bottom two panels of Figure~\ref{fig:DLA_los} over $8400~{\mbox \AA} \lesssim \lambda \lesssim 8500~{\mbox \AA}$).  As noted above, however, in this work we conservatively consider only the 1000~\kms\ region within which \meandiff\ is measured (although see Section~\ref{sec:ionization}).  We thus conclude that damping wings from compact, high-column density absorbers can plausibly produce features at the edges of proximity zones similar to the \lya\ feature in the J0148 spectrum.

\subsection{Neutral islands}\label{sec:island}

\begin{figure*}
   \centering
   \begin{minipage}{\textwidth}
   \begin{center}
   \includegraphics[width=0.95\textwidth]{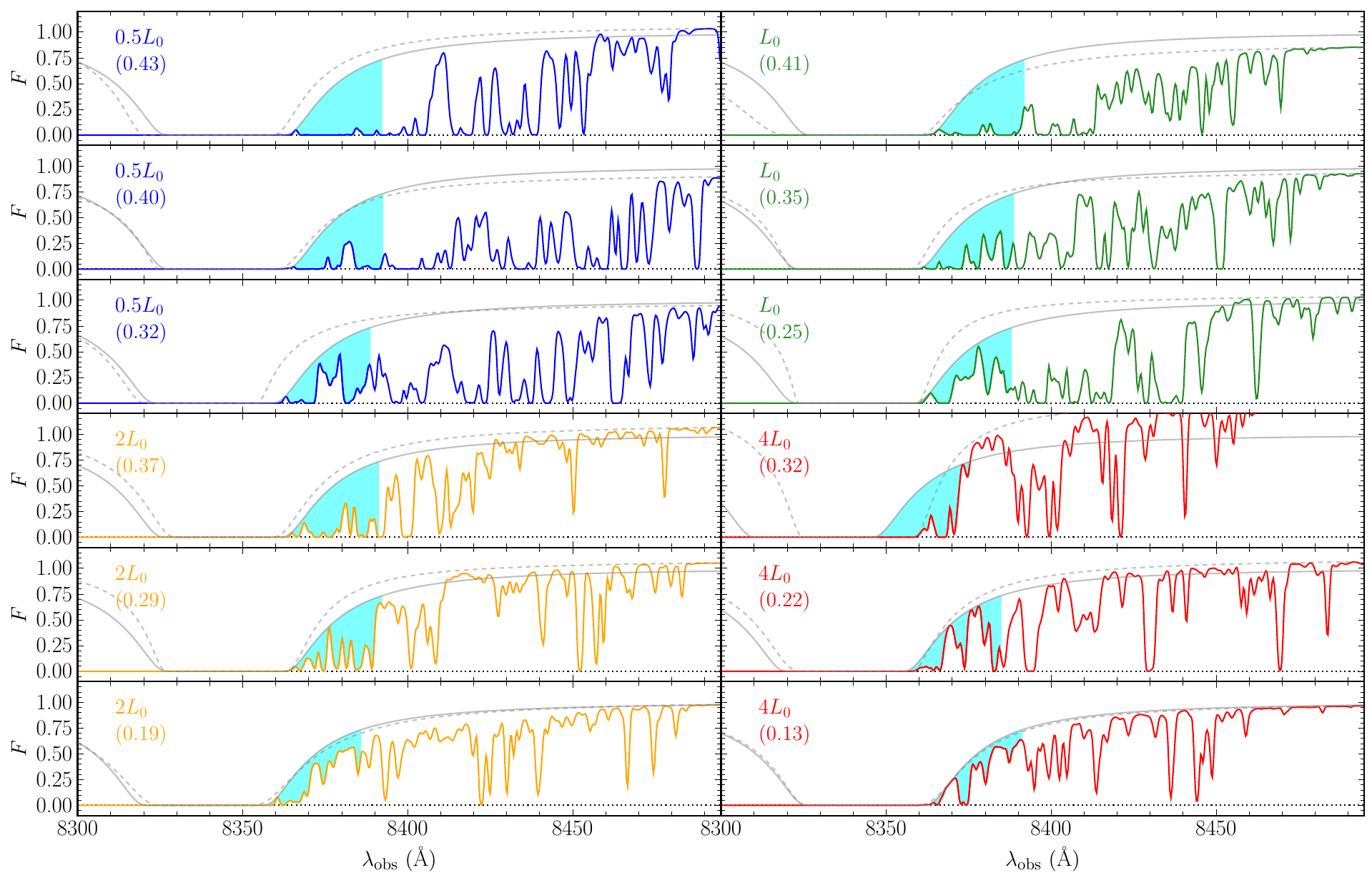}
   \vspace{-0.05in}
   \caption{Simliar to Figure~\ref{fig:LLS_los}, but for simulated lines of sight ending in 7.5~\hinvMpc\ neutral islands.  The dashed line in each panel shows the actual damping profile for the neutral island, which has been rescaled by the same amount as the transmitted flux.  The smooth solid line is a template profile for a  mean-density 7.5~\hinvMpc\ neutral island that has been shifted to produce the minimum allowable \meandiff\ over the 1000 \kms\ shaded region at the red edge of the template damping wing.}
   \label{fig:island_los}
   \end{center}
   \end{minipage}
\end{figure*}

\begin{figure}
   \begin{center}
   \includegraphics[width=0.43\textwidth]{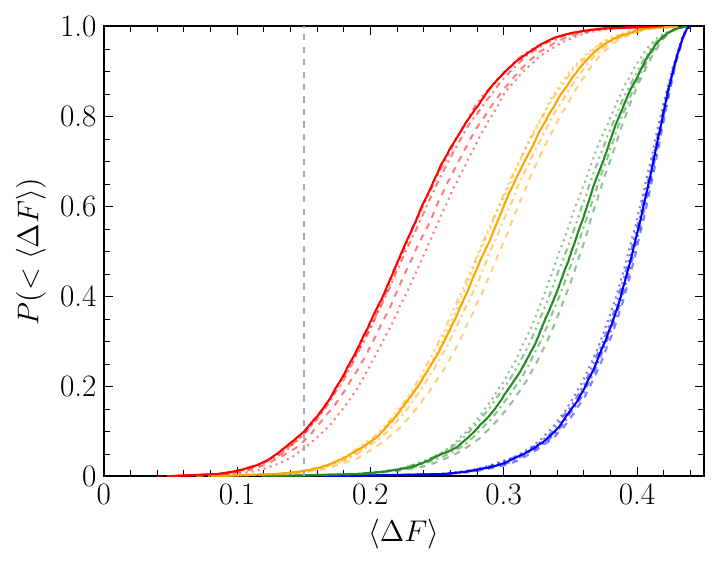}
   \vspace{-0.05in}
   \caption{Similar to Figure~\ref{fig:LLS_stats}, but for simulated proximity zones ending in neutral islands.  Solid lines are for the fiducial island length of 7.5~\hinvMpc.  Other line styles are for lengths 2.5 (dotted), 5.0 (dash-double-dotted), 10 (dash-dotted), and 15 (dashed) \hinvMpc.  Colors correspond to the quasar luminosity scalings in Figure~\ref{fig:prox_scaling}.  The left-to-right ordering of the line groups is $L_{\rm q} = 4L_0, 2L_0, L_0, 0.5L_0$.  The vertical dashed line shows \meandiff\ for J0148.}
   \label{fig:island_stats}
   \end{center}
\end{figure}

Finally, we consider whether the smooth profile at the edge of the J0148 proximity zone could be caused by the damping wing from an extended neutral island.  Here we generate artificial proximity zones as described in Section~\ref{sec:LLS}, but set $x_{\rm H\,I} = 1$ over a 7.5 \hinvMpc\ interval starting 34 \hinvMpc\ from the quasar.  The choices of length and distance are motivated by the largest mean-density neutral island allowed by the observed \lya\ and \lyb\ transmission at the end of the J0148 proximity zone (see Section~\ref{sec:J0148}).  We note that in terms of \meandiff, the distance will be largely degenerate with the \Lquasar\ scaling.

Example simulated proximity zones ending in neutral islands are shown in Figure~\ref{fig:island_los}.  Similar to Figure~\ref{fig:LLS_los}, lines of sight at the 5$^{\rm th}$, 50$^{\rm th}$, and 95$^{\rm th}$ percentiles in \meandiff\ are shown for each \Lquasar\ scaling.  The full \meandiff\ distributions are plotted in Figure~\ref{fig:island_stats}.  The results are broadly similar to those for DLAs in that the higher \Lquasar\ values can produce significant percentages of lines of sight with \meandiff\ values at or below the J0148 value (1\% for $2L_0$, 10\% for $4L_0$).  In these cases (e.g, bottom two panels of Figure~\ref{fig:island_los}), the template damping profile used to measure \meandiff\ (solid line) is reasonably well aligned with the actual damping wing produced by the neutral island along the line of sight (dashed line).  The nearness of the tops of the transmission peaks to the dashed line again suggests that the transmission would otherwise be high within the \meandiff\ region, a point we return to below.  Here we simply conclude that the damping wing profiles created by large neutral islands can plausibly produce features with \meandiff\ values similar to J0148.

We can also test whether the consistency with J0148 depends sensitively on the length of the neutral island.  To check this, we generate mock proximity zones ending in islands with lengths 2.5, 5.0, 10, and 15~\hinvMpc.  In each case the red end of the island is 34~\hinvMpc\ from the quasar.  The \meandiff\ distributions are shown in Figure~\ref{fig:island_stats}, and we note that we are still calculating \meandiff\ using the 7.5~\hinvMpc\ mean-density template.  Although there is some sensitivity to the length of the island, the changes with respect to our fiducial length of 7.5~\hinvMpc\ are modest.  We therefore conclude that while the damping wing from a neutral island can produce a \meandiff\ value similar to the J0148 value, the agreement is not strongly dependent on the island length.

\section{Discussion: origin of the J0148 damping wing}\label{sec:discussion}

\subsection{Compact absorber vs. neutral island}

\begin{figure}
   \begin{center}
   \includegraphics[width=0.40\textwidth]{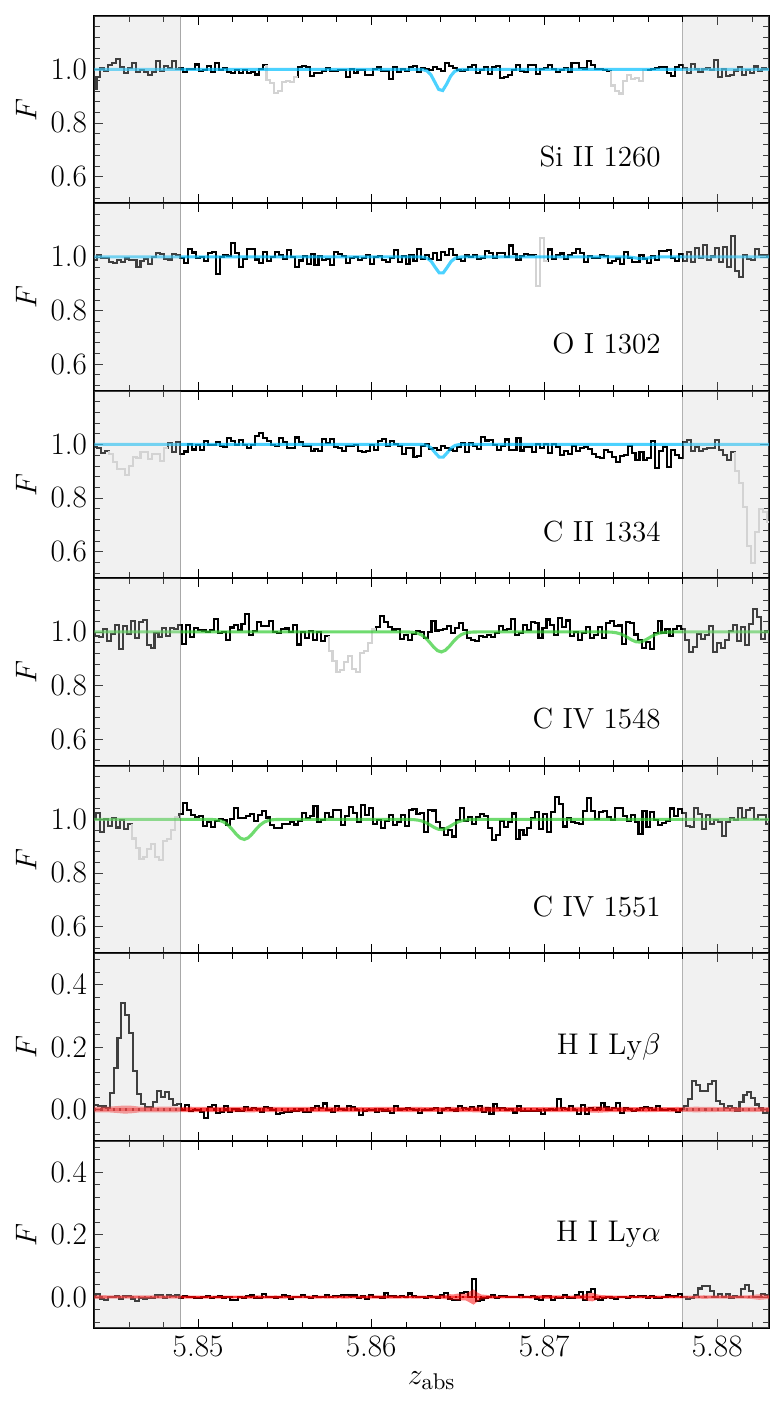}
   \vspace{-0.05in}
   \caption{Illustration of limits on metal lines at redshifts over the \lya+\lyb\ trough at the end of the J0148 proximity zone.  Each panel shows the normalized flux over the wavelengths corresponding to the indicated ion over absorption redshifts $5.844 < z_{\rm abs} < 5.883$.  Unrelated absorption lines are plotted in light grey.  Shaded regions at the edges contain significant \lya\ and/or \lyb\ transmission.  The blue line in the top three panels shows the expected absorption for a low-ionization absorber with an \oi\ column density of $\log{(N_{\rm O\, I} / {\rm cm^{-2}})} = 13.2$ and relative abundances similar to known $z \simeq 6$ absorbers (see text for details).  Green lines in the next two panels show the \civ\ transmission for $\log{(N_{\rm C\, IV} / {\rm cm^{-2}})} = 12.8$.  Note the vertical scale in the upper five panels.  Red shaded regions in the lower two panels show the 1$\sigma$ flux uncertainty.}
   \label{fig:metals}
   \end{center}
\end{figure}

The above analysis demonstrates that a smooth transmission feature at the end of proximity zone similar to the J0148 \lya\ transmission window strongly requires a damping wing.  The damping wing can in principle be created either by a compact absorber, such as a DLA, or an extended neutral island.  Here we argue, however, that while both scenarios are allowed in terms of their \meandiff\ statistics, a neutral island is favored based on other factors.

There are at least three reasons to favor an extended neutral island.  The first is the lack of any corresponding metal absorption lines.  As noted by \citet{becker2019} and \citet{davies2023}, the J0148 spectrum exhibits no metal absorption systems at $z > 5.5$.  In particular, no metal absorption is seen in either low-ionization (e.g., \ion{O}{i}, \ion{C}{ii}, \ion{Si}{ii}, \ion{Mg}{ii}) or high-ionization (\ion{C}{iv}, \ion{Si}{iv}) lines over the redshifts of the \lya+\lyb\ trough at the end of the proximity zone ($z \simeq  5.85$--5.88; Figure~\ref{fig:metals}).  The deep X-Shooter data excludes systems with $\log{(N_{\rm O\, I} / {\rm cm^{-2}})} \gtrsim 13.2$ for a Doppler $b$-parameter of 10~\kms.  In Appendix~\ref{sec:compact}, we show that a compact absorber would need an \hi\ column density of at least $\sim$$10^{19.5}~{\rm cm^{-2}}$ in order for the probability of generating a \meandiff\ value equal to or less than the J0148 values to be at least 1\%, even for the highest \Lquasar\ scaling we consider.  Assuming that oxygen is in charge exchange equilibrium with hydrogen, these limits on \oi\ and \hi\  corresponds to a maximum oxygen metallicity of ${\rm [O/H]} \lesssim -3.0$,\footnote{Using photospheric solar abundances from \citet{asplund2009}, and assuming no dust depletion.} which is comparable to the most metal-poor DLA known \citep{cooke2017}.  We show the expected line profiles for such an absorber in Figure~\ref{fig:metals}, where the column densities for \ion{Si}{ii} and \ion{C}{ii} are scaled with respect to \oi\ based on the relative abundances measured by \citet{becker2012} for low-ionization absorbers near $z = 6$ (see also Sodini et al., in prep).   The non-detection of \siii~$\lambda$1260 would increase the metallicity constraint to ${\rm [O/H]} \lesssim -3.1$ for the same abundance ratios.  The presence of such an ultra-low-metallicity, low-ionization absorber is clearly ruled out at these redshifts.  We also show the expected \civ\ absorption for $\log{(N_{\rm C\, IV} / {\rm cm^{-2}})} = 12.8$ and $b = 30$~\kms, which is also excluded by the data.  If a compact absorber is creating the damping wing observed at the edge of the J0148 proximity zone then it must be extremely metal poor.  We note, however, that some DLAs at these redshifts may have metallicities near ${\rm [O/H]} \simeq -3$ \citep[e.g.,][]{dodorico2018,banados2019}, so the lack of metal lines on its own does not fully rule out a compact absorber.

A second reason to favor a neutral island is that the evidence of a damping wing is found adjacent to an extremely long and opaque giant \lya\ trough.  In our simulated proximity zones we set the \lya\ transmission to zero blueward of the proximity zone by artificially increasing the \hi\ neutral fraction.  In more physically motivated models, however, it has been shown that such long troughs are most likely to occur when a line of sight intersects one or more neutral islands \citep[e.g.,][]{kulkarni2019,keating2020a,nasir2020}.  In a companion paper, Zhu et al. (in prep) show that long dark troughs over $5 \lesssim z \lesssim 6$ are statistically associated with damping wing absorption, which is consistent with a model in which the troughs arise from neutral islands.  

Finally, a deep search for [\oiii]-emitting galaxies with the {\it James Webb Space Telescope} as part of the EIGER survey reveals no sources within the potential DLA redshift window of $5.85 \lesssim z \lesssim 5.88$ \citep{eilers2024}.  Based on the lack of metal lines, the presence of an extreme \lya\ trough, and the lack of any directly detected galaxy, we therefore conclude that the damping wing over the J0148 \lya\ transmission window is more likely to arise from an extended neutral island than a compact absorber.

\subsection{Physical characteristics of the high-transmission window}\label{sec:charachteristics}

\begin{figure}
   \begin{center}
   \includegraphics[width=0.46\textwidth]{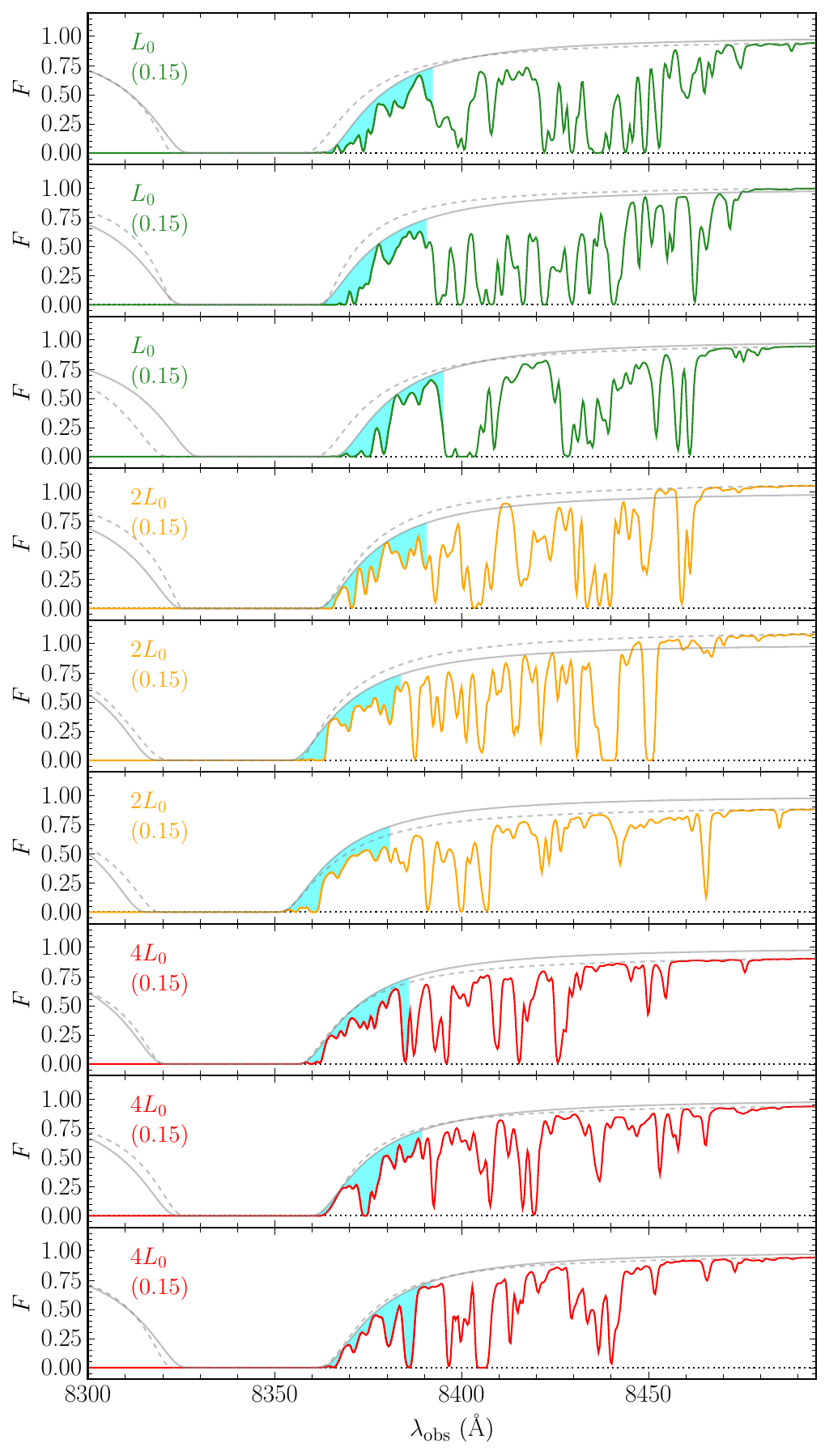}
   \vspace{-0.05in}
   \caption{Simliar to Figure~\ref{fig:island_los}, but for simulated lines of sight ending in 7.5~\hinvMpc\ neutral islands that have \meandiff\ values similar to the J0148 value.  Three lines of sight are shown for quasar luminosity scalings $L_{\rm q} = L_0$ (top), $2L_0$ (middle), and $4L_0$ (bottom).  Lines of sight similar to J0148 are extremely rare in these simple models for $L_{\rm q} = L_0$, while they are more common for $2L_0$ and $4L_0$ (Figure~\ref{fig:island_stats}).}
   \label{fig:similar_los}
   \end{center}
\end{figure}

We now turn to examining the physical conditions that may give rise to transmission features such the J0148 window.  For illustration, in Figure~\ref{fig:similar_los} we plot a sample of simulated lines of sight with \meandiff\ similar to the J0148 value for $L_{\rm q} = L_0$, $2L_0$, $4L_0$.  In many cases the tops of the transmission peaks within the \meandiff\ region nearly reach the damping wing envelope produced by the neutral island.  (This is somewhat less true for $L_{\rm q} = L_0$ than for $2L_0$ and $4L_0$, but cases similar to J0148 for $L_{\rm q} = L_0$ are extremely rare; see Figure~\ref{fig:island_stats}.)  This suggests that the regions adjacent to these neutral islands are highly transmissive in \lya, which can be achieved via high ionization rates, low densities, and/or elevated gas temperatures.  We consider each of these factors in the following sections

\subsubsection{Ionization rate}\label{sec:ionization}

It is clear from Figures~\ref{fig:DLA_stats} and \ref{fig:island_stats} that the probability of producing a line of sight with \meandiff\ equal to or less than the J0148 value increases towards higher quasar luminosities.  The higher ionization rates produce lower neutral fractions (Figure~\ref{fig:prox_scaling}),  and hence more transmission against which the damping wing profile becomes apparent.  In our models the ionization rate within the proximity zone is determined by a combination of the quasar luminosity and ionizing opacity, and for simplicity we only vary \Lquasar\ (Figure~\ref{fig:prox_scaling}).  While this approach may be adequate to bracket the range of possible ionization rates, invoking a high quasar luminosity to produce a window of high transmission may be inconsistent with other characteristics of the J0148 proximity zone.  Notably, if we select the simulated lines of sight with 
$\langle \Delta F \rangle \le 0.2$ (cf. the J0148 value of 0.15) and compute their mean transmission in a 2000~\kms\ region from 8400~\AA\ (typically just reward of the \meandiff\ region) to 8456~\AA\ (10 \hinvMpc\ from the quasar), all of the lines of sight have higher  transmission than J0148, even when random continuum errors are included.  \citet{keating2015} find that the properties of $z \sim 6$ quasar proximity zones only weakly depend on the host halo mass.  The fact that our lines of sight do not start at massive halos is therefore unlikely to account for this difference in transmission (see also Section~\ref{sec:generating}).  We also note that our composite-based quasar continuum (Figure~\ref{fig:J0148_spec}) is highly consistent with the UV spectrum we adopt to compute the ionizing emissivity (Section~\ref{sec:generating}).  This may make a factor of two or more departure from the nominal ionizing luminosity unlikely, although quasar variability could be playing a role \citep[e.g.,][]{davies2020,satyavolu2023}.

This inconsistency redward of the \meandiff\ region, particularly with the $L_{\rm q} = 2L_0$ and $4L_0$ models, which produce the largest transmission, suggests that there may be other factors driving the high transmission of the J0148 window.  One possibility is that there are local sources contributing to the ionization rate near the end of the J0148 proximity zone.  Indeed, \citet{eilers2024} identify a group of [\oiii]-emitting galaxies at $z \simeq 5.92$, coincident with the strong \lya\ absorption just redward of the J0148 window.  In order for the probability of producing a \meandiff\ value consistent with J0148 to exceed 1\%, similar to the $2L_0$ case in Figure~\ref{fig:island_stats}, such sources would need to roughly triple the ionization rate of our baseline $L _{\rm q} = L_0$ case,\footnote{The difference in the ionization rate between the $L_0$ and $2L_0$ cases includes an additional factor of 1.5 due to the difference in ionizing opacity within the proximity zone; see Section~\ref{sec:generating}.} which is already seven times the background rate at the center of the transmission window at $z=5.897$ ($\lambda_{\rm obs} = 8385$~\AA; Figure~\ref{fig:prox_scaling}).  Other ways of boosting the J0148 window transmission include low densities and/or high gas temperatures, which we examine below.

\subsubsection{Density}

\begin{figure}
   \begin{center}
   \includegraphics[width=0.46\textwidth]{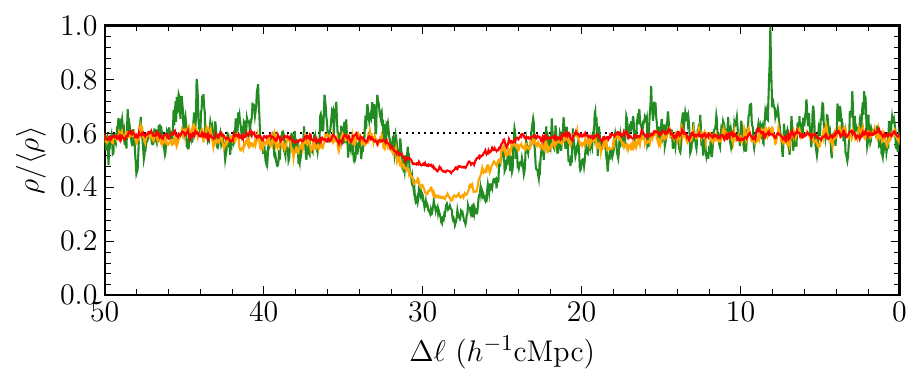}
   \vspace{-0.05in}
   \caption{Median line-of-sight overdensities for lines of sight ending in 7.5~\hinvMpc\ neutral islands with \meandiff\ values within $\pm$0.05 of the J0148 value.  Colors correspond to the quasar luminosity scaling values in Figure~\ref{fig:prox_scaling}, although only $L_{\rm q} = L_0$ (green), $2L_0$ (orange), and $4L_0$ (red) are shown here.  These lines of sight tend to contain an underdensity near the end of the quasar proximity zone, with a median depth that increases with decreasing quasar luminosity.}
   \label{fig:density}
   \end{center}
\end{figure}

The role of density can be examined using the line-of-sight gas densities for our mock proximity zones.  We select lines of sight with neutral islands that have \meandiff\ values within $\pm$0.05 of the J0148 value (i.e., 0.10 to 0.20) for $L_{\rm q} = L_0$, $2L_0$, $4L_0$.  We then compute the median density along these lines of sight separately for each $L_{\rm q}$.  The results are shown in Figure~\ref{fig:density}.  The median profiles show clear underdensities from $\sim$24 to $\sim$32~\hinvMpc\ from the quasar.  This closely corresponds to the position of the J0148 transmission window, and is roughly adjacent to the near edge of the simulated neutral islands.  The depth of the median underdensity increases with decreasing \Lquasar, suggesting that the role of density is especially important when the local ionization rate is lower.  The underdensities required for the $L_{\rm q} = L_0$ case are extremely rare, however, which is why only $\sim$0.1\% of these lines of sight have \meandiff\ values equal to or less than the J0148 value of 0.15, and only $\sim$1\% have $\langle \Delta F \rangle \le 0.2$.

\subsubsection{Temperature}

\begin{figure}
   \begin{center}
   \includegraphics[width=0.43\textwidth]{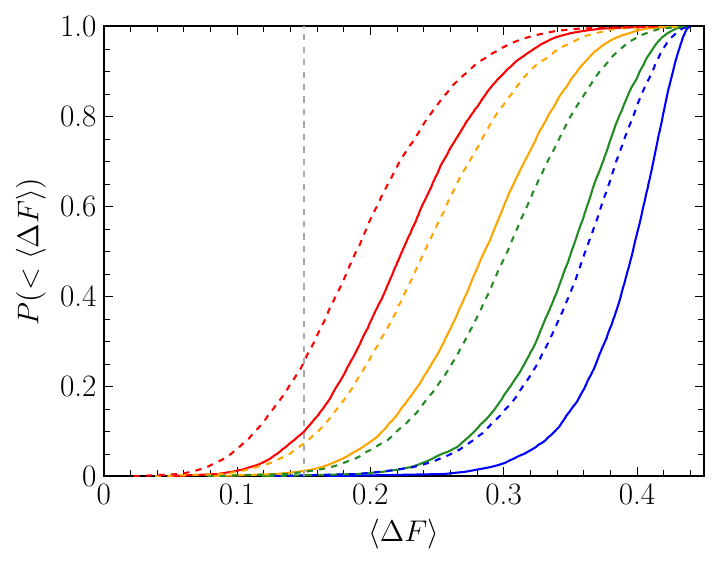}
   \vspace{-0.05in}
   \caption{Cumulative probability distributions of \meandiff\ for simulated lines of sight ending in neutral islands.  The solid lines are the same as those plotted in Figure~\ref{fig:island_stats}.  The dashed lines are for the same lines of sight with the gas temperature at mean density increased from 12,000~K to 30,000~K.  The left-to-right ordering of both line styles is $L_{\rm q} = 4L_0$ (red), $2L_0$ (orange), $L_0$ (green), $0.5L_0$ (blue).  In terms of the change in \meandiff, increasing the gas temperature by this amount is roughly equivalent to doubling the quasar luminosity.}
   \label{fig:Tboost_stats}
   \end{center}
\end{figure}

Finally, we consider the role of gas temperature in producing transmission features similar to the J0148 window.  In the preceding models we adopted a temperature-density relation with a temperature at mean density of 12,000 K (see Section~\ref{sec:generating}).  The temperature may be higher locally, however, if the gas was recently reionized.  We tested the impact of elevated temperatures on the \meandiff\ distribution by recalculating the optical depths for simulated lines of sight with neutral islands after increasing the gas temperature to a uniform value.  The choice of temperature is motivated by radiative transfer calculations from \citet{daloisio2019}, which predict post-ionization temperatures as a function of the ionization front velocity and the spectral index of the incident ionizing radiation.  Calculating the ionization front velocity as $v_{\rm ion} = \dot{n}_{\rm ion} / n_{\rm H\, I}$, where $\dot{n}_{\rm ion}$ is the incident flux of ionizing photons and $n_{\rm H\, I}$ is the neutral hydrogen number density, a front driven by a quasar with $L_{\rm q} = L_0$ would be moving at $\sim$25,000~\kms\ through mean-density gas once it reached the edge of the J0148 proximity zone.  This includes an attenuation of the quasar flux by $\sim$$1/e$ due to opacity within the proximity zone predicted by the model described in Section~\ref{sec:generating}.   \citet{daloisio2019} find a post-I-front temperature of $\sim$30,000 K at this velocity for our adopted ionizing spectral index of $\alpha = 1.5$.  The incident spectrum may be somewhat harder due to spectral filtering within the proximity zone, though the \citet{daloisio2019} relations suggest that this would not strongly impact the temperature.  The temperature could be as high as $\sim$40,000 K if helium is doubly ionized by the quasar \citep{bolton2012,davies2016a}.
For simplicity, we increase the temperature to 30,000 K along the entire line of sight, but again note that we are evaluating \meandiff\ only within a narrow region at the end of the proximity zone.  

The resulting \meandiff\ distributions are plotted in Figure~\ref{fig:Tboost_stats}.  Increasing the temperature to 30,000 K substantially increases the number of lines of sight with \meandiff\ values equal to or less than J0148, roughly comparable to the effect of doubling the quasar luminosity.  We note that most of the change in the \meanflux\ distribution comes from the lower neutral fractions associated with suppressed recombination rates rather than from increased thermal broadening. 

\subsection{A recently reionized region?}

In the preceding sections we have shown that a smooth \lya\ transmission window such as the one towards J0148 is more likely to occur if the absorber producing the damping wing is adjacent to a region of the IGM that has a high ionization rate, low density, and/or elevated temperature.  Here we develop a possible explanation of the transmission window motivated by these factors and the nearness of the window to both an extended \lya+\lyb\ absorption trough and a bright quasar.

An association between neutral islands and low-density regions may occur naturally near the end of reionization.  Low-density regions are expected to be among the last regions of the IGM to be ionized due to the scarcity of ionizing sources.  Indeed, long \lya\ troughs such as the one towards J0148 have been shown to trace low-density regions in terms of the number density of nearby galaxies \citep{becker2018,kashino2020,christenson2021}.   
The edge of the J0148 proximity zone, which is adjacent to a deep \lya+\lyb\ trough, may be part of such a low-density region.  

The boundaries of neutral islands will also be the sites of ionization fronts, and hence recently heated gas.  In order for photoionization heating to increase the contrast in transmission between the edge of the proximity zone and the interior, the interior would need to have been reionized significantly earlier, giving it time to cool.  This might have occurred if the quasar turned on after the interior was already  ionized (although we note that if it was ionized by soft spectra then the quasar might still have doubly ionize helium in the interior, decreasing the temperature contrast with the edge).
We can estimate whether J0148 would be able to drive an ionization front at the edge of its proximity zone using a simple Str\"{o}mgren sphere model.  Adopting the nominal ionizing luminosity parameters from Section~\ref{sec:generating}, assuming mean density gas, and ignoring recombinations, producing an ionized region out to 34~\hinvMpc\ would require a quasar lifetime of $\sim$30~Myr.  Including the radially varying ionizing opacity described in Section~\ref{sec:generating} would increase the required time by roughly a factor of two, although somewhat less time would be required if part of the proximity zone was already ionized when J0148 began its luminous phase, or if the region near the edge of the proximity zone is underdense.  A lifetime of tens of Myr is consistent with, albeit at the upper end of recent estimates of the lifetime of typical $z \sim 6$ quasars based on the sizes of their proximity zones \citep{durovcikova2024}.  It is therefore plausible that J0148 could be driving an ionization front at the end of its proximity zone.  There may also be additional sources contributing to the ionizing flux.   As noted above, \citet{eilers2024} find a group of [\oiii]-emitting galaxies just redward of the J0148 transmission window, as well as an overdensity of [\oiii] emitters associated with the quasar itself.

The recent passage of an ionization front through a low-density region thus provides a possible explanation for the J0148 transmission window.  Driven by ionizing photons from the quasar and/or nearby sources, the front would have passed through the edge of a neutral island, leaving behind hot  gas.  Such gas would be highly transparent to \lya\ photons, producing a strong transmission window that is then modified by the damping wing from the foreground neutral island.  This scenario is consistent with the unusual strength and shape of the transmission window, its nearness to a giant absorption trough, and the proximity of a bright quasar.  As noted above, the high transmission of the J0148 window may also be produced by an extremely low-density region and/or enhanced ionization from nearby sources.

\section{Summary}\label{sec:summary}

We have presented evidence that the longest and deepest known \lya\ trough at $z < 6$ is associated with damping wing absorption.  We find that the  profile of a strong \lya\ transmission window at the end of the J0148 proximity zone, adjacent to the red end of trough, requires a damping wing and is highly unlikely to arise from resonant absorption alone.  The lack of either metal lines or directly detected galaxies further suggests that the damping wing does not arise from a compact absorber such as a DLA.  Instead, the we find that the damping absorption is more likely to arise from an extended neutral island, particularly given the close physical association between the proximity zone transmission window and the extended \lya+\lyb\ trough.

The characteristics of the J0148 transmission window suggest that it is associated with a region of  intrinsically low \lya\ opacity.  It may be a deep void, a region where the local ionizing background is enhanced by local sources in addition to the quasar, and/or a region with elevated gas temperatures.  One possibility is that the window traces hot gas that has been recently ionized by the quasar and/or other nearby sources.  

This work has focused on a somewhat unique line of sight, and our analysis has been purposefully narrow in addressing the question of whether the characteristics of an individual transmission feature require the presence of a damping wing.  We expect that a more comprehensive and robust analysis of $z \lesssim 6$ quasar proximity zones, particularly those that end in extended troughs, may provide further evidence of neutral islands.  In a companion paper, Zhu et al. (in prep) find a statistical association between  \lya+\lyb\ troughs and damping wing absorption.  These works are the first direct evidence that extended \lya+\lyb\ troughs at $z < 6$ arise from islands of neutral gas, and hence that hydrogen reionization is still ongoing at these redshifts.

\section*{Acknowledgements}

We thank the referee, Fred Davies, for helpful comments and suggestions.  We also thank Sarah Bosman for providing the J0148 [\cii] redshift ahead of publication, as well as Anson D'Aloisio and Chris Cain for helpful discussions.  G.D.B., Y.Z., and S.H. were supported by the NSF through grant AST-1751404.  Y.Z. was also supported by the NSF through award SOSPADA-029 from the NRAO. J.S.B. was supported by STFC consolidated grants ST/T000171/1 and ST/X000982/1.

This work is based in part on observations made with ESO Telescopes at the La Silla Paranal Observatory under program ID 084.A-0390.

The hydrodynamical simulations used in this work were performed with supercomputer time awarded by the Partnership for Advanced Computing in Europe (PRACE) 8th Call. We acknowledge PRACE for awarding us access to the Curie supercomputer, based in France at the Tres Grand Centre de Calcul (TGCC).  This work has also used the DiRAC Data Analytic system at the University of Cambridge, operated by the University of Cambridge High Performance Computing Service on behalf of the STFC DiRAC HPC Facility (www.dirac.ac.uk). This equipment was funded by BIS National E-infrastructure capital grant (ST/K001590/1), STFC capital grants ST/H008861/1 and ST/H00887X/1, and STFC DiRAC Operations grant ST/K00333X/1. DiRAC is part of the National E-Infrastructure.

\section*{Data Availability}

The raw data used this article are available from the ESO archive at http://archive.eso.org/cms.html.  A reduced version of the J0148 X-Shooter spectrum is available through a public github repository of the XQR-30 spectra at https://github.com/XQR-30/Spectra.

\bibliographystyle{mnras}
\bibliography{damping_wing_refs}

\appendix

\section{quasar continuum estimation}\label{sec:continuum}

Here we provide further details on the continuum estimation technique used to normalize the J0148 spectrum.  The continuum was constructed from a composite of lower-redshift quasar spectra drawn from the Sloan Digital Sky Survey Data Release 16 (DR16) quasar catalog \citep{lyke2020}.  We use spectra with $S/N > 5$ per pixel that are not flagged as broad absorption line objects (BALs).  We also use only wavelength redward of 3800~\AA, avoiding bluer wavelengths where the noise in DR16 spectra tends to increase.  As we describe below, the spectra are chosen based on their similarity to J0148 over wavelengths redward of the \lya\ forest.  They are then corrected for intervening \lya\ and \lyb\ absorption and averaged to create an estimate of the unabsorbed continuum.  

In many respects our approach is similar to other matched composite techniques in the literature, especially those where \civ\ emission is used to select the matching spectra \citep[e.g.,][]{mortlock2011,simcoe2012,bosman2015}.  A key difference with these and other approaches, however, is that we statistically correct for foreground \lya\ and \lyb\ absorption.  Provided that there are enough objects contributing to the composite to largely overcome the scatter in IGM transmission between lines of sight, this approach, which we describe bellow, allows us to avoid the potentially complicating step of fitting continua over the forest of the lower-redshift spectra \citep[for a review of continuum reconstruction techniques, see][]{bosman2022}.

Our first step is to fit a spline to the J0148 spectrum redward of the \lya\ forest.  This provides a basis for comparison that minimizes the impact of metal absorption lines and noise.  We note that J0148 exhibits extended \civ\ broad absorption (Figure~\ref{fig:J0148_spec}) that obscures the continuum  between the \siiv\ and \civ\ emission lines.  This part of the spectrum is not used to select DR16 spectra, however, and thus the presence of the BAL does not impact our continuum estimate.  We also note that the red edge of the BAL falls $\sim$20,000~\kms\ blueward of the systemic redshift.  Any corresponding broad absorption in \nv\ or \lya\ would therefore fall blueward of the proximity zone and not impact our \meandiff\ measurement.

The DR16 spectra are treated flexibly in two ways.  First, we allow the redshifts to vary.  This accounts for redshift errors in the DR16 catalog (as well as for J0148, in principle, although it has an accurate systemic redshift from [\cii] 158 $\mu$m emission; see Bosman et al., in prep).  It also provides a means of accommodating velocity shifts of the emission lines with respect to the systemic redshift.  This is particularly helpful for J0148, which has a \civ\ emission line blueshift of $\sim$3000~\kms.  Allowing the redshifts to vary provides a pool of potential matches that is significantly larger than what would be available if the DR16 spectra were fixed at their (potentially inaccurate) catalog redshifts.  The results of this approach are tested below.  For each object, we start by adopting a redshift that places the peak of the \civ\ emission line at the same rest-frame wavelength as the J0148 \civ\ peak.  We then allow the DR16 spectrum to shift in velocity by up to $\pm$1000~\kms\ in increments of 250~\kms.  Second, we adjust the overall slope and scaling of the the DR16 spectra to match J0148.  We do this by multiplying the DR16 spectrum by a power law in wavelength.  At each velocity offset, the power law is fit to the median fluxes over rest-frame wavelengths 1270--1390~\AA\ and 1600--1750~\AA.

Our primary goal is to estimate the continuum over the J0148 transmission window near $\lambda \simeq 8370$--8400~\AA, which falls at the blue edge of the \lya\ emission line.    
The shifted and scaled DR16 spectra are thus compared to J0148 over the two wavelength regions redward of the forest that best capture the variation in the strength and shape of \lya: the red side of the \lya+\nv\ complex (1216--12151~\AA\ rest-frame) and the \civ\ emission line (1500--1600~\AA\ rest-frame). These regions are highlighted in Figure~\ref{fig:J0148_spec}.  The inclusion of \civ\ is motivated by the known correlation between \civ\ and \lya\ emission \citep[e.g.,][]{richards2011,greig2017,davies2018}.  We note that by including the red side of \lya+\nv\ we assume that there is no significant damping wing absorption redward of 1216~\AA.  For J0148 this is probably a reasonable assumption given the extent of the proximity zone and the lack of any obvious DLAs within it  (Figure~\ref{fig:J0148_prox}).  We compute the median values of the J0148 spline and the individual DR16 spectra in bins of 7~\AA\ (rest-frame) over these regions, where the binning is chosen to minimize the impact of noise and absorption lines in the DR16 spectrum while preserving the overall spectral shape.  The goodness-of-fit is determined from the mean squared difference between the binned values.  We create composites of the 200 best-fitting spectra, where the number is chosen to balance the accuracy of fits to the emission lines with the suppression of scatter over the forest.

Before computing the composite, each individual spectrum is corrected for foreground \lya\ and \lyb\ absorption.  This is done by diving each pixel in the \lya\ and \lyb\ forests by the expected mean transmission at the corresponding redshift, taking into account the quasar proximity effect.  The baseline mean \lya\ transmission is taken from \citet{becker2013a}.  Baseline \lyb\ transmission values are scaled from the \lya\ values using a hydrodynamical simulation of the IGM.  Specifically, we use outputs from the 40-2048 run of the Sherwood simulation suite \citep{bolton2017}.  We rescale the ionization rates to match the mean \lya\ transmission at a given redshift, and then measure the mean \lyb\ transmission from the resulting \lyb\ optical depths.  The proximity effect is taken into account using a model similar to the one described in \citet{becker2021}.  For each object, we compute the distance, $R_{\rm eq}$, at which the ionization rate from the quasar would be equal to the background ionization rate in the absence of attenuation or redshifting of the quasar photons.  Following \citet{becker2021}, we compute the ionizing luminosity of the quasar based on its absolute magnitude at rest-frame 1450~\AA, assuming a broken power-law of the form $L \propto \nu^{-\alpha}$ with $\alpha = 0.6$ over $912~{\mbox \AA} < \lambda_{\rm rest} < 1450~{\mbox \AA}$, and $\alpha = 1.5$ at $\lambda_{\rm rest} < 912$~\AA\ \citep[see][]{lusso2015}.  Background ionization rates are interpolated from values in \citet{becker2013} over $2.4 < z < 4.0$ and \citet{bolton2005} at $z = 2.0$.  Using this value of $R_{\rm eq}$, we calculate the expected ratio of the total (quasar $+$ background) ionization rate to the background rate as a function of distance from the quasar.  The expected mean \lya\ and \lyb\ transmission at each distance is then computed after rescaling the ionization rate in the simulation by this ratio.

We note that this method does not take into account the fact that quasars will tend to live in overdensities.  It will therefore tend to under-correct the absorption close to the quasar redshift, and hence leave a flux deficit in the composite near the peak of the \lya\ emission line.  We find, however, that this effect is significantly reduced when using DR16 spectra at lower redshifts, where associated absorbers stand out more strongly and can be rejected more easily (see below).  We therefore generate composites from two samples.  The first composite, $F_{{\rm hi}z}$, is drawn from $\sim$16,000 spectra over $2.91 < z < 4.0$.  The upper redshift bound is chosen to limit the correction for foreground absorption, while the lower bound is chosen to provide coverage down to Ly$\gamma$ at observed wavelengths redward of 3800~\AA.  The second composite, $F_{{\rm low}z}$, is drawn from $\sim$32,000 spectra over $2.2 < z < 2.5$, where the redshift range is intended to reduce the impact of associated absorption.  The two composites for J0148 are generally similar to within $\sim$5\% over the \lya\ forest, except near the peak of \lya, where $F_{{\rm hi}z}$ has a $\sim$12\% deficit with respect to $F_{{\rm low}z}$.  The final composite is an average of the two at wavelengths where they each have coverage from at least 100 individual spectra, apart from 1195~\AA\ to 1216~\AA\ in the rest-frame (the \lya\ peak) where only $F_{{\rm low}z}$ is used.  The composite shown in Figure~\ref{fig:J0148_spec} uses $F_{{\rm hi}z}$ at $\lambda < 7909$~\AA, $F_{{\rm low}z}$ over $8353~{\mbox \AA} < \lambda < 8499~{\mbox \AA}$ (the \lya\ peak), and an average of the two at other wavelengths.

We apply a modest rejection scheme when calculating the mean composite values.  At a given rest-frame wavelength, individual pixels are rejected if they are more than 3$\sigma$ outliers from the median.  They are also rejected if they are more than 3$\sigma$ outliers after smoothing the individual spectra by a Gaussian kernel with ${\rm FWHM} = 1000$~\kms.  The first criterion is mainly meant to reject sharp features such as bad pixels or strong metal lines, while the second criterion is meant to exclude extended features such as DLAs.  We note, however, that in the forest the scatter includes the sightline-to-sightline variation in the \lya\ and \lyb\ absorption, making unwanted features more difficult to detect and the rejection relatively mild.  

Finally, the uncertainty in the composite continuum is estimated directly from the scatter among the individual spectra.  We compute the ratio of the individual fluxes (corrected for \lya\ and \lyb\ transmission) to the composite flux within a sliding window of 5000~\kms.  The 1$\sigma$ uncertainty in the composite is taken to be the standard deviation in this ratio.  We note that this approach is conservative in that it includes both the scatter in the intrinsic quasar continua and the scatter in the \lya\ and \lyb\ transmission between sightlines, which will be non-negligible even on 5000~\kms\ scales.

\begin{figure*}
   \centering
   \begin{minipage}{\textwidth}
   \begin{center}
   \includegraphics[width=1.0\textwidth]{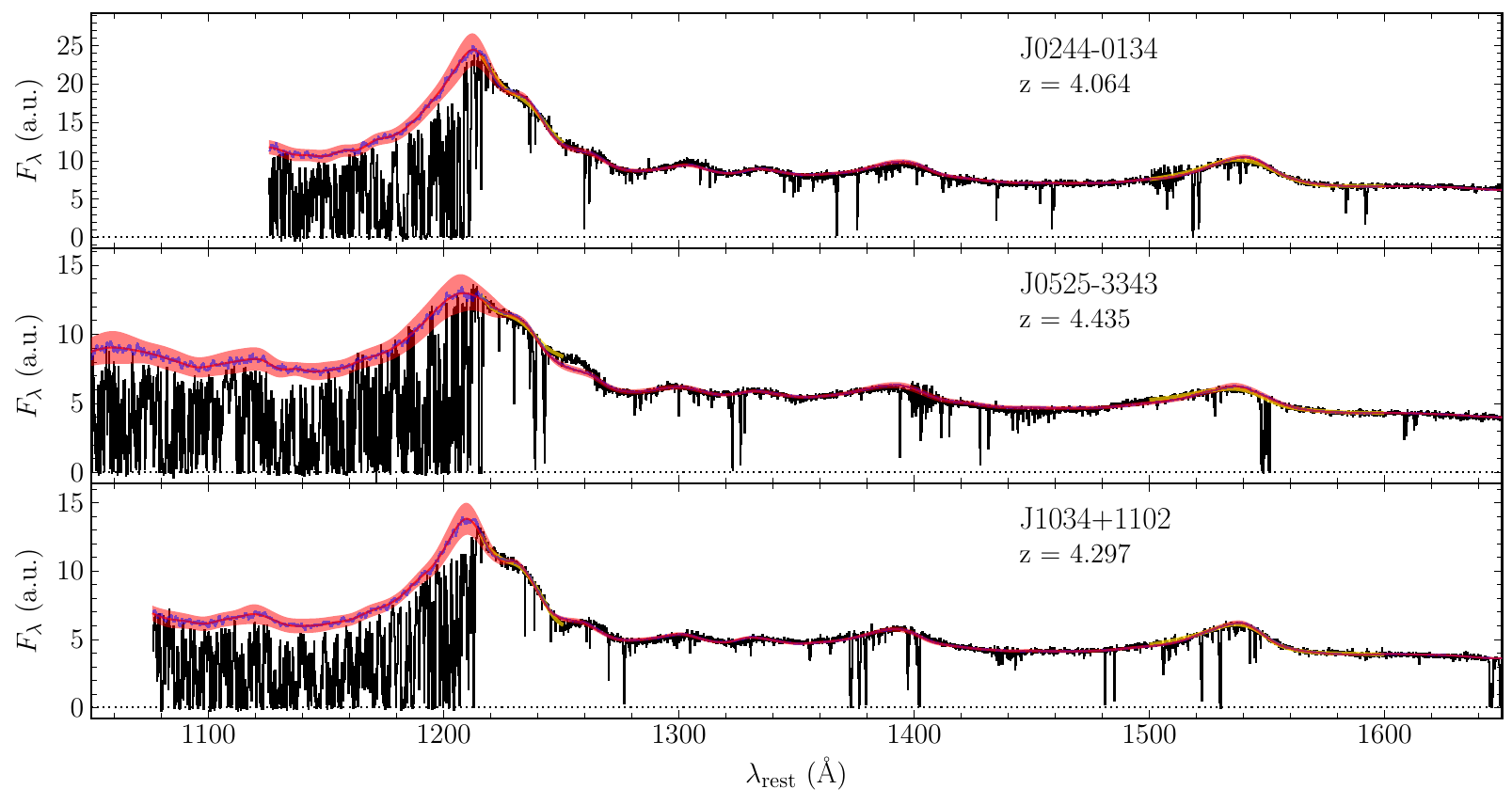}
   \vspace{-0.05in}
   \caption{Example continuum fits to three quasars at $z \sim 4$ drawn from the XQ-100 sample.  The spectra are from the VIS arm of X-Shooter.  Similar to Figure~\ref{fig:J0148_spec}, the two wavelength regions to which the DR16 spectra are matched are shown with yellow lines, which are largely obscured by the composites.  The raw composites are plotted as blue histograms.  Spline fits to the raw composite are shown with a smooth red line, while the 1$\sigma$ uncertainty in the composite is given by the shaded region.  Note that the composites generally trace the tops of the transmission peaks over the \lya\ forest.}
   \label{fig:example_continua}
   \end{center}
   \end{minipage}
\end{figure*}

We test our continuum estimation technique by applying it to lower-redshift objects.  At $z \lesssim 4$, the intrinsic continuum over the \lya\ and \lyb\ forests can be roughly inferred from the peaks of the transmission features, providing a check on the composite results.  Motivated by our focus on J0148, we choose three $z \sim 4$ objects for illustration with strong \civ\ blueshifts that have high-quality X-Shooter spectra from the XQ-100 sample \citep{lopez2016} : J0244$-$0134 ($z_{\rm eq} = 4.064$), J0525$-$3343 ($z_{\rm em} = 4.435$), and J1034$+$1102 ($z_{\rm eq} = 4.297$).  The fits to these objects are shown in Figure~\ref{fig:example_continua}.  We note that we are plotting spectra from the VIS arm of X-Shooter only in order to avoid uncertainties in the scaling between the UVB and VIS arms.  Nevertheless, the continuum estimates roughly trace the tops of the \lya\ forest transmission peaks.  For this work we have not attempted to evaluate the accuracy of our technique for a broad range of quasar spectral shapes; however, the results for these objects provide some confidence that composite shown in Figure~\ref{fig:J0148_spec} provides a reasonable estimate of the unabsorbed J0148 continuum.

\section{Numerical convergence}\label{sec:convergence}

Our simulated lines of sight are drawn from the 80-2048 run from the Sherwood simulation suite \citep{bolton2017}.  This uses an 80~\hinvMpc\ box with $2 \times 2048^3$ particles, giving a gas particle mass of $\sim$$8 \times 10^5~h^{-1}~{\rm M}_\odot$.  To check whether our results depend on either box size or resolution, we repeat our analysis of simulated proximity zones ending in LLSs (Section~\ref{sec:LLS}) using the 80-1024, 40-2048 and 40-1024 runs from the same suite.  The 40-1024 run has the same mass resolution as the 80-2048 run but is a factor of eight smaller in volume.  The 80-2048 and 80-1024 runs, meanwhile, have the same volume but differ in mass resolution by a factor of eight.  The 40-2048 and 40-1024 differ in mass resolution by the same factor.

\begin{figure}
   \begin{center}
   \includegraphics[width=0.43\textwidth]{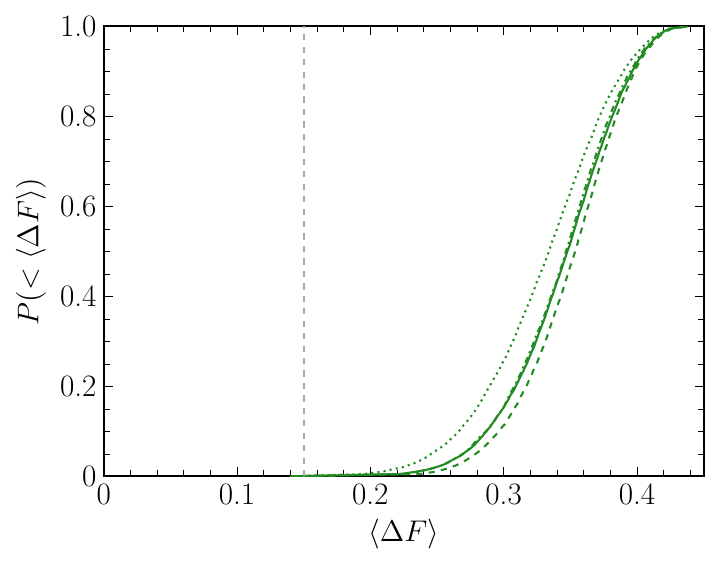}
   \vspace{-0.05in}
   \caption{Results of numerical convergence tests.  The four lines show the \meandiff\ distributions for simulated proximity zones ending in LLSs for a luminosity scaling of $L = L_0$.  The solid line is for a 80-2048 run and is the same as in Figure~\ref{fig:LLS_stats}.  The dotted, dashed, and dash-dotted lines are for the 80-1024, 40-2048 and 40-1024 runs, respectively.}
   \label{fig:numerical}
   \end{center}
\end{figure}

The results for our nominal quasar luminosity scaling ($L = L_0$) are shown in Figure~\ref{fig:numerical}.  The 80-2048 and 40-1024 runs produce nearly identical results, suggesting that the 80-2048 run is not missing a substantial number of rare voids that would increase the number of lines of sight with low \meandiff\ values.  The higher resolution of the 80-2048 run compared to the 80-1024 run pushes the \meanflux\ distribution to slightly higher values, largely because taller, sharper transmission peaks prevent the \meanflux\ region of the damping wing template from overlapping as much of the proximity zone transmission (see Section~\ref{sec:LLS}).  A similar though somewhat smaller shift is seen between the 40-1024 and 40-2048 runs.  The results suggests that further increasing the resolution would tend to disfavor the LLS model even more strongly, although the effect would be modest.  We conclude that numerical effects are unlikely to have a significant impact on our results.

\section{High column-density compact absorbers}\label{sec:compact}

\begin{figure}
   \begin{center}
   \includegraphics[width=0.43\textwidth]{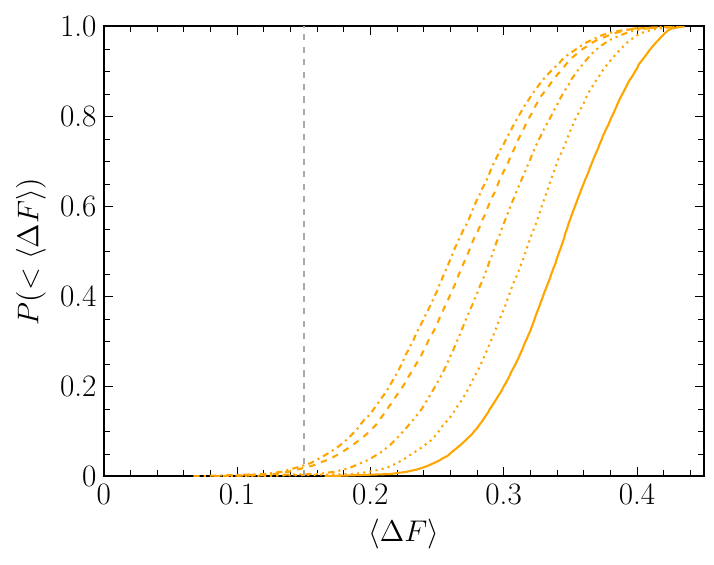}
   \vspace{-0.05in}
   \caption{Similar to Figure~\ref{fig:LLS_stats}, but for simulated proximity zones ending in compact absorbers with a range of column densities.  Results shown here are for luminosity scaling $L = 2L_0$.  As a reference, the solid line is for LLSs with column density $\log{(N_{\rm H\, I} / {\rm cm^{-2}})} = 18.0$, and is the same as the one plotted in Figure~\ref{fig:LLS_stats}.  Other lines are for \hi\ column densities of  $\log{(N_{\rm H\, I} / {\rm cm^{-2}})} = 19.0$ (dot), 19.5 (dash-dot-dot), 20 (dash-dot), and 20.5 (dash)} 
   \label{fig:NHI_stats}
   \end{center}
\end{figure}

\begin{figure}
   \begin{center}
   \includegraphics[width=0.46\textwidth]{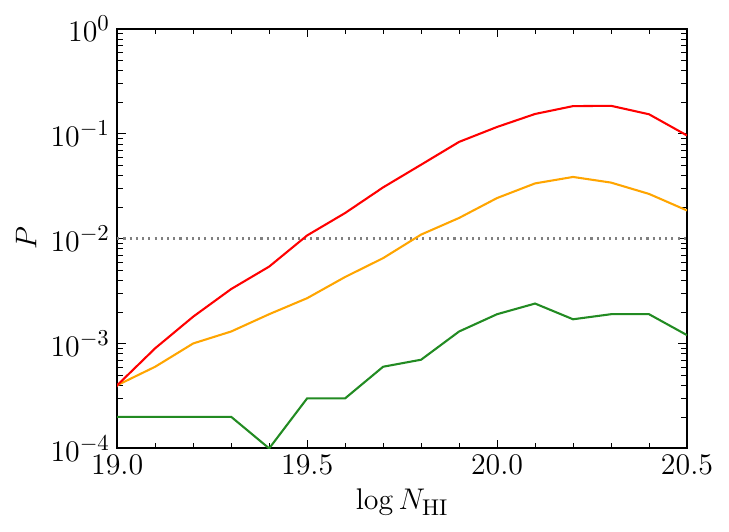}
   \vspace{-0.05in}
   \caption{Probability that \meandiff\ is equal to or less than the J0148 value for simulated proximity zones ending in high-column density compact absorbers, as a function of  \hi\ column density.  The lower, middle, and upper solid lines are for $L_{\rm q} = L_0$, $2L_0$, and $4L_0$, respectively.  The probability tends to peak between $\log{(N_{\rm H\, I} / {\rm cm^{-2}})} = 20.0$ and 20.5.  A horizontal line at $P = 0.01$ is plotted for reference.}
   \label{fig:NHI_prob}
   \end{center}
\end{figure}

In Section~\ref{sec:DLA} we focused on compact absorbers with the canonical DLA \hi\ column density of $\log{(N_{\rm H\, I} / {\rm cm^{-2}})} = 20.3$.  Here we provide results for a wider range of column densities.  Figure~\ref{fig:NHI_stats} shows \meandiff\ distributions for compact absorbers with column densities ranging from $\log{(N_{\rm H\, I} / {\rm cm^{-2}})} = 19.0$ to 20.5, along with the Lyman limit case (i.e., no damping wing) from Section~\ref{sec:LLS}.  For clarity, we plot the $L_{\rm q} = 2L_0$ scaling only.  The \meandiff\ distribution shifts towards smaller values with increasing $N_{\rm H\, I}$ up to $\log{(N_{\rm H\, I} / {\rm cm^{-2}})} = 20.0$, but then shifts back towards higher values at $\log{(N_{\rm H\, I} / {\rm cm^{-2}})} = 20.5$.  This behavior reflects the change in the compact absorber damping wing profile with column density.  Figure~\ref{fig:NHI_prob} plots the probability that a line of sight will have \meandiff\ equal to or less than the J0148 value as a function of $N_{\rm H\, I}$.  Here we plot the results for $L = L_0$, $2L_0$, and $4L_0$;  $0.5L_0$ is not shown because it does not produce a significant number of lines of sight with small enough \meandiff\ values.  Even $L_{\rm q}$ requires $\log{(N_{\rm H\, I} / {\rm cm^{-2}})} \ge 19.5$ in order for the probability to exceed 1\%.  In each case the probability peaks near $\log{(N_{\rm H\, I} / {\rm cm^{-2}})} \simeq 20.2$.  Our focus on the canonical DLA column density of $\log{(N_{\rm H\, I} / {\rm cm^{-2}})} = 20.3$ is thus a conservative choice in that it roughly maximizes the probability that a compact absorber will produce a transmission feature that is similar to J0148 in terms of \meandiff.

\bsp	
\label{lastpage}
\end{document}